\documentclass[12pt,a4paper]{article}
\usepackage{jheppub}
\usepackage{amsmath}
\usepackage{amsfonts}
\usepackage{amssymb}
\usepackage{graphicx}
\usepackage{mathrsfs}
\usepackage{comment}
\usepackage{bbold}

\newcommand{\e}{\textrm{e}}
\newcommand{\nCr}[2]{\begin{pmatrix}#1 \\ #2\end{pmatrix}}
\newcommand{\PBC}{\textrm{PBC}}

\newcommand{\TBC}{\textrm{TBC}}
\newcommand{\tr}{\textrm{tr}}
\newcommand{\hx}{\hat{x}}

\centerline{\title{\boldmath $U(N)$ Yang-Mills in non-commutative space time.}}

\author[a]{Naser Ahmadiniaz,}
\author[b,c]{Olindo Corradini,}
\author[d]{James P. Edwards,}
\author[e]{Pablo Pisani}

\affiliation[a]{Center for Relativistic Laser Science, Institute for Basic Science, 61005 Gwangju, Korea}
\affiliation[b]{Dipartimento di Scienze Fisiche, Informatiche e Matematiche,
 Universit\`a degli Studi di Modena e Reggio Emilia, Via Campi 213/A, I-41125 Modena, Italy}
\affiliation[c]{INFN, Sezione di Bologna, Via Irnerio 46, I-40126 Bologna, Italy}
\affiliation[d]{Instituto de F\'isica y Matem\'aticas
Universidad Michoacana de San Nicol\'as de Hidalgo
Edificio C-3, Apdo. Postal 2-82
C.P. 58040, Morelia, Michoac\'an, M\'exico}
\affiliation[e]{Instituto de F\'{i}sica La Plata, CONICET-UNLP
C.C. 67 (1900) La Plata, Argentina}
\emailAdd{ahmadiniaz@ibs.re.kr}
\emailAdd{olindo.corradini@unimore.it}
\emailAdd{jpedwards@cantab.net}
\emailAdd{pisani@fisica.unlp.edu.ar}

\abstract{We present an approach to $U_\star(N)$ Yang-Mills theory in non-commutative space based upon a novel phase-space analysis of the dynamical fields with additional auxiliary variables that generate Lorentz structure and colour degrees of freedom. To illustrate this formalism we compute the quadratic terms in the effective action focusing on the planar divergences so as to extract the $\beta$-function for the Yang-Mills coupling constant. Nonetheless the method presented is general and can be applied to calculate the effective action at arbitrary order of expansion in the coupling constant, including both planar and non-planar contributions, and is well suited to the computation of low energy one-loop scattering amplitudes.}

\begin{document}
\maketitle

\section{Introduction}

In the present manuscript our focus is on advancing some new calculational methods for non-commutative gauge field theories. These gauge field theories are defined in a space-time endowed with a non-commutative product, which we take to be the familiar Moyal product. Such theories become non-local when this Moyal product is applied to fields~\cite{Douglas:2001ba, Szabo:2001kg, ComplScal} and appear naturally in some quantum mechanical models and string theory \cite{String1, NCS1, String2, String3}. This feature of non-locality, along with the presence of an intrinsic minimal length scale, make  non-commutative field theories mimick some aspects of theories of quantum gravity \cite{Doplicher:1994tu}, which is one of the main reasons why they still draw quite some attention in the literature. Due to the presence of so-called UV/IR mixing \cite{Minwalla:1999px}, which involves non-planar diagrams and may affect the (non)renormalisability of such theories,  of particular interest are the renormalisation properties of non-commutative gauge field theories, and thus of their one-loop effective actions. This is a prototypical example of a study which can be performed using first-quantised, or \textit{worldline}, techniques. Indeed, it has been shown that non-commutative field theories are linked to phase-space worldline path integrals. This was developped in \cite{Bonezzi:2012vr, SatoNC} for (complex-) scalar field theories and in \cite{NCU1, KiemKim} for a $U(1)$ gauge theory. 

The worldline formalism was initially developped as an efficient tool to obtain one-loop QCD amplitudes and effective actions~\cite{Bern:1990cu, Strassler:1992zr}, but has by now become a well established approach to performing numerous computations in various Quantum Field Theories (QFT)~\cite{ChrisRev}. Originally inspired by string theory, it provides a first quantised approach to QFT that in many cases demonstrates superior calculational efficiency and better manifests gauge invariance. In the area of non-Abelian QFTs some examples where first quantised approaches have been used as the main working tool  include strong-coupling expansions~\cite{Halpern:1977ru, Halpern:1977he}, multi-loop effective actions~\cite{Sato:1998sf}, the spinning particle approach to Yang Mills amplitudes~\cite{Dai:2008bh, Reuter:1996zm, ColTree, JO1, JO2}, as well as to the Standard Model physics~\cite{Mansfield:2014vea} and to Grand Unified scenarios~\cite{Edwards:2014bfa}. 

In this work we develop a phase-space worldline model to study the one-loop effective action of $U(N)$ Yang-Mills (YM) theories on $D$-dimensional (Euclidean) non-commutative spacetime that provides a completely first-quantised representation of the one-loop effective action of a non-Abelian vector field, combining and extending recent advances in the worldline description of colour and spin degrees of freedom \cite{JO1, JO2, CreteProc}. In this way, the additional information about both colour and spin degrees of freedom will appear as path integrals over auxiliary variables. A significant advantage of this approach is that it is no longer necessary to carry out the path ordering usually required to deal with matrix valued potentials that enter the worldline action since this prescription is produced automatically by integration over the auxiliary variables. 

Moreover, we shall demonstrate that our approach unifies contributions from the ``gauge sector'' (associated to the physical degrees of freedom of the vector field) and the ``ghost sector'' (that maintains gauge invariance) into a single model -- selection of each sector is achieved by simply fixing a Chern Simons level in a path integral over further auxiliary variables. Thus the extended worldline model seems to be a useful and general approach to describing all aspects of our calculations in a democratic manner. From this final worldline representation we extract the $U_{\star}(N)$ $\beta$-function from the field theory two-point function as a simple application and verification of our model. 

Following this philosophy, our main result is that, upon applying the background field method, the one-loop effective action for a $U(N)$ Yang-Mills field on Moyal space-time can be written as a difference of two terms that represent the contribution from the gauge sector and the ghost sector respectively, \textit{viz.} 
\begin{align}
	\Gamma[A]= \Gamma[A]_{1,1} - 2\Gamma[A]_{1, 0}
\end{align}
where
\begin{align}
\Gamma[A]_{\varrho, r} :&=-\frac{1}{2} \int_{0}^{\infty}\frac{d T}{(4 \pi)^{\frac{D}{2}}T^{1 + \frac{D}{2}}} \int \mathscr{D}p \oint \mathscr{D}x \oint\mathscr{D}[\bar{\psi}, \psi] \oint \mathscr{D}[\bar{c}, c]  \int_{0}^{2\pi} \frac{d \varphi}{2\pi}\,  \int_0^{2\pi} \frac{d\vartheta}{2\pi} \nonumber\\&\times \exp{\left(-S_{\varrho, r}[p, x, \bar{\psi}, \psi, \bar{c}, c, \varphi, \vartheta]\right)}.
\end{align}
The integral over $T$ is the Schwinger proper time integral familiar in worldline calculations, whilst $p$ and $x$ represent the momentum and coordinates of the embedding of periodic point particle trajectories in (commutative) Euclidean space $\mathbb{R}^D$. Following this, the conjugate pairs $\bar{\psi}$and $\psi$ and $\bar{c}$ and $c$ are auxiliary Grassmann worldline fields that produce the Hilbert spaces associated to the spin and colour degrees of freedom. Finally, the modular integrals over $\varphi$ and $\vartheta$ are what remain after gauge fixing local unitary symmetries associated to the spin and colour fields. These integrals act to project onto the subspace of the Hilbert space associated to spin $r$ forms and the fully anti-symmetric $U(N)$ representation with dimension $\varrho$ through Chern-Simons terms in the worldline action
\begin{align}
	&S_{\varrho, r}[p, x,  \bar{\psi}, \psi,\bar{c}, c, \varphi, \vartheta]= \int_{0}^{1}dt\bigg[-i p \cdot \dot{x} + p^{2} + \bar{c}^{a}\dot{c}_{a} + \bar{\psi} \cdot \dot{\psi} + T V_{a}{}^{b}\left(\bar{c}^{a} c_{b} + \frac{1}{2}\delta^{a}_{b}\right)+ \nonumber \\[2mm]
	\hspace{-2em}&- 2iT	 \bar{\psi}^{\mu}(\mathcal{F}_{\mu \nu})_{a}{}^{b}\psi^{\nu}\left(\bar{c}^{a} c_{b} + \frac{1}{2}\delta^{a}_{b}\right)+ i\vartheta\left(\bar{c}^{a}c_{b} -  n\right) + i\varphi \left(\bar{\psi} \cdot \psi - s\right)\bigg].
\end{align}
Here $n \equiv \varrho - \frac{N^{2}}{2}$ and $s \equiv r - \frac{D}{2}$ are the Chern-Simons levels that are all that is needed to distinguish the different sectors, $V[A]_{a}{}^{b}$ is a potential representing the orbital coupling of the vector field and $(\mathcal{F}[A]_{\mu\nu})_{a}{}^{b}$ a potential for the spin coupling, both of whose forms we shall give below. The auxiliary fields remove the need to path order the exponential in $\Gamma[A]_{\varrho, r}$, as we shall explain in section \ref{AWF}. 

To illustrate our method we expand the effective action to quadratic order in the vector field, which is sufficient to determine the $\beta$-function of the theory. Since the divergences of the non-planar diagrams are interpreted as IR divergences, they do not affect the behaviour of the $\beta$-function, which is solely determined by UV-divergent planar diagrams; for this reason we disregard non-planar contributions and content ourselves with the computation of planar ones. The divergent contribution to the change in the bare action from the sum of planar contributions from the gauge sector and the ghost sector (in $D = 4$, with $\frac{1}{2g^{2}}$ denoting the bare coupling and with $\Lambda$, $m$ UV- and IR-cut-offs) is
\begin{equation}
	\delta S = -\frac{1}{2g^2}\int_{\mathbb{R}^4}dx \, \frac{11 N}{48 \pi^{2}} g^{2}\log \left( \frac{\Lambda^2}{m^2}\right)
  \left\{A_\mu^a(x)\left(-\delta_{\mu\nu}\,\partial^2+\partial_\mu\partial_\nu\right)A_\nu^a(x)\right\}
\end{equation}
from which follows the $\beta$-function from the running of the coupling constant:
\begin{equation}
	\beta(g) = -\frac{11 N}{48 \pi^{2}}g^{3}.
\end{equation}
This is in agreement with previous calculations \cite{Krajewski:1999ja,Martin:1999aq} and verifies our worldline approach yields the correct result. However we argue that the unification of both sectors into a single worldline theory that enables an efficient determination of the Feynman diagrams contributing at each order is a notable step in the development of this approach to QFT. 

We organise the manuscript as follows. In Section~\ref{EA} we use the background field method to extract from the YM action the quadratic operators for the YM fields and for the Faddeev-Popov ghosts following from a gauge fixing. The functional integrals over the quantum fields yield the one-loop effective action, which we represent in terms of a heat-trace and thus in terms of a phase-space worldline path integral. In Section~\ref{AWF} we introduce some auxiliary worldline fields which are helpful in the perturbative expansion of the effective action: their quantisation reproduces the desired representation of the gauge group and of the Lorentz group, and avoids the needs for a path ordering prescription. In the remaining parts of the manuscript we use the phase-space particle model to compute one- and two-point functions and evaluate the one-loop $\beta$-function of the theory. We end this paper with conclusions and an outlook in Section \ref{discussion}. In Appendix \ref{app1} we collect some useful group identities involving the structure constants and Appendix \ref{pbc} contains the derivation of the worldline Green functions for gauge and ghost fields and some of their expectation values for completeness.   


\section{Yang-Mills on non-commutative space-time}
\label{EA}

Here we define the Moyal space-time and the $U_\star(N)$ YM theory that we study on this space. We will consequently calculate its one-loop effective action using the background field method.

\subsection{Moyal space-time}

In a $D$-dimensional Euclidean Moyal space time, the usual space-time coordinates, $x^{\mu}$, are replaced by operators, $\hat{x}^{\mu}$, for which the fundamental commutator is introduced, 
\begin{equation}
	\left[\hx^\mu, \hx^\nu\right] = i\theta^{\mu\nu},
	\label{CommC}
\end{equation}
where the constant, real, skew-symmetric matrix $\theta^{\mu\nu}$ represents the non-commutativity parameters \cite{Douglas:2001ba, Szabo:2001kg}. This relation presents itself in various areas of quantum mechanics and string theory \cite{String1,String2}.

There is a well-known correspondence between the non-commuting operators, $\hat{x}^{\mu}$, and an algebra of functions on $\mathbb{R}^{D}$ under a non-commuting product exhibited by the Weyl symbol
\begin{align}
	\hat{\mathcal{O}}\left[f\right] &= \int_{\mathbb{R}^D} dx\, f(x)\hat{\Delta}\left(x\right); \nonumber \\
	 \hat{\Delta}(x) &= \int_{\mathbb{R}^D} \frac{dk}{\left(2\pi\right)^{D}}\ e^{i k \cdot \hx} e^{-i k \cdot x}.
\end{align}
This map can be used to define the Moyal $\star$-product \cite{Moyal,Moyal2} from the product of two Weyl operators evaluated in position space
\begin{align}
	\hat{\mathcal{O}}\left[f\right]\hat{\mathcal{O}}\left[g\right] &= \hat{\mathcal{O}}\left[f\star g\right]; \nonumber \\
	f(x)\star g(x) &= f(x) \exp{\left(\frac{i}{2}\overset{\leftarrow}{\partial_{\mu}}\,\theta^{\mu\nu}\!\stackrel{\rightarrow}{\partial_{\nu}}\right)}g(x).
	\label{star}
\end{align}
This product is just a deformation of the familiar algebra of functions \cite{Product} on $\mathbb{R}^{n}$ (for discussion of non-commutative space described in polar coordinates see \cite{andy, borel, NCBTZ}) and makes use of a common notation where derivatives act in the direction of their overhead arrows. In Fourier space the Moyal product of a number of functions of the operators $\hat{x}_{i}$ reads\footnote{In this notation $\mathcal F(\phi)=\tilde \phi$, $\bar \delta=(2\pi)^D\delta$ and $d\bar p=dp/(2\pi)^D$.}
\begin{align}
  \mathcal{F}(\phi_1\star \phi_2\star \ldots)(p)&=\int \prod_i d\bar{p}_i
  \ \bar\delta\big(\textstyle{\sum_i\,}p_i-p\big)
  \ \tilde \phi_1(p_1)\,\tilde \phi_2(p_2)\ldots\ e^{-i\sum_{i<j}p_i \cdot \theta  \cdot p_j}\,.
\end{align}
This allows us to define a field theory action for pure Yang-Mills theory on Moyal space-time based upon ordinary fields but with a deformed product.

Non-commutativity in momentum space is encompassed in the twisting factor $e^{-i\sum_{i<j}p_i \cdot \theta  \cdot p_j}$. This non-polynomial combination of incoming momenta at the vertex is responsible for the non-locality of the interaction. Notice also that the twisting factor is absent in the Moyal product of only two fields; for this reason, non-commutativity has no effect on quadratic expressions so it does not affect propagators but rather only interaction terms. This property also implies that the Moyal product is invariant under cyclic permutation of the fields, which considerably reduces the number of interaction terms in the construction of non-commutative generalisations of commutative theories.

\subsection{Yang-Mills theory}

To define our theory we introduce a one-form, $A_{\mu} = A_{\mu}^{a} T^{a}$, valued in the fundamental representation of the Lie algebra $u(N)$ with generators $\{T_{a}\}$, and its field strength tensor
\begin{align}
  F_{\mu\nu}&=i\,[D_\mu,D_\nu]_{\star}=\partial_\mu A_\nu-\partial_\nu A_\mu-i\,[A_\mu,A_\nu]_{\star}
  \label{Fstar}
\end{align}
where $D_\mu=\partial_\mu-iA_\mu$ is the covariant derivative and $\star$ indicates that the Moyal product is to be used for multiplication. Using\footnote{From here one can read off the defining (anti-)commutators of the Lie algebra\begin{align}
	[T^{a}, T^{b}] &= if^{abc}T^{c}\,, \ a=0,\dots, N^2-1 \\
	\{T^{a}, T^{b}\} &= d^{abc}T^{c}
	\end{align}
	where the structure constants $f^{abc}$ are completely anti-symmetric whilst the $d^{abc}$ are fully symmetric. The ``time-like'' component is the $U(1)$ generator $T^0=\frac{1}{\sqrt{2N}} {\mathbb 1}$, and we have  $f^{ab0}=0$ and $d^{ab0}=\delta^{ab}\sqrt{2/N}$. }
\begin{align}
T^a T^b =\frac12 (if^{abc} +d^{abc})T^c~,
\end{align} 
the components of the field-strength tensor read
\begin{align}
  F^a_{\mu\nu}&=\partial_\mu A^a_\nu-\partial_\nu A^a_\mu
  -\frac{i}{2}\,d^{abc}\,[A^b_\mu,A^c_\nu]_{\star}+\frac{1}{2}\,f^{abc}\,\{A^b_\mu,A^c_\nu\}_{\star}\,.
\end{align}
This field strength tensor transforms covariantly under the gauge field transformation
\begin{align}
  A_\mu(x)\rightarrow U\star A_\mu\star U^{-1}+i\,U\star \partial_\mu U^{-1}\,,
  \label{Atrans}
\end{align}
where
\begin{align}
  U=e_\star^{i\alpha(x)}=1+i\alpha(x)-\tfrac12\,\alpha(x)\star \alpha(x)+\ldots
\end{align}
and $\alpha(x)=\alpha^a(x)\,T^a$ is Lie-algebra valued. 

An action that is invariant under the symmetry transformation (\ref{Atrans}) is
\begin{align}
S[A]=\frac{1}{2g^2}\int_{\mathbb{R}^D}dx\,{\rm tr}\,( F_{\mu\nu}\star F_{\mu\nu})
\end{align}
where $g^2$ is the bare coupling constant, and the trace is computed in the fundamental representation of the gauge group. Note that the field strength tensor given in (\ref{Fstar}) means that the anti-commutator of the gauge group algebra enters the action alongside the commutator familiar in commutative space. This means that we require the Lie algebra of the generators in the chosen representation to close under anti-commutation, which is not the case for $SU(N)$; the simplest extension compatible with this requirement is the $U(N)$ group which we study for the remainder of this article.

\subsection{Effective Action}
Since this quantity will occupy our calculations in the following, we briefly recap the definition and determination of the effective action in QFT. For a general quantum field $\Phi(x)$ described by a classical action $S[\Phi]$, the effective action $\Gamma_{\rm eff}[\Phi]$ is defined as
\begin{equation}
\e^{-	\Gamma_{\rm eff}[\bar\Phi] } = \int \, \mathscr{D}\Phi \, \e^{-S[\bar\Phi+\Phi]+\int \bar J(x)\Phi(x)}\,,
\label{GammaDef}
\end{equation}
where $\bar J(x)$ is the source that generates the arbitrary background field $\bar\Phi$, that is, $ \bar J(x)=\delta_{\bar\Phi}\Gamma_{\rm eff}$. Expanding the previous expression around the background field $\bar\Phi$ and recalling that $\bar{\Phi}$ is the expectation value in the presence of the source $\bar{J}$, one obtains, to leading order in quantum corrections,
\begin{equation}
\e^{-	\Gamma_{\rm eff}[\bar\Phi] } =\e^{-	S[\bar\Phi] } \int \, \mathscr{D}\Phi \, \e^{-\frac12\int \Phi\,\delta_{\bar\Phi}^2S\,\Phi\, + \, \ldots}\,.
\end{equation}
Therefore, the effective action can be written as
\begin{align}
	\Gamma_{\rm eff}[\bar\Phi]=S[\bar\Phi]+\Gamma[\bar\Phi]+O(\hbar^2)\,,
\end{align}
where the one-loop corrections, $\Gamma[\bar\Phi]$, to the effective action are given by
\begin{align}\label{effact}
	\Gamma[\bar\Phi]=\pm\frac12\log{\rm Det}
	\left\{\frac{\delta^2 S[\bar\Phi]}{\delta\bar\Phi(x)\delta\bar\Phi(x')}\right\}\,.
\end{align}
The field $\Phi$ is used to collectively denote a set of fields (real, complex conjugates etc.), eventually comprising Grassmann components. The upper sign in \eqref{effact} corresponds to the contribution of each bosonic field in $\Phi$; the lower, to each Grassmann component.

Next, we shall calculate the one-loop corrections to the effective action $\Gamma[A]$ for the non-commutative Yang-Mills field by using the background field method, expressing the resulting functional determinants in terms of heat-traces and passing to the first-quantised (worldline) representation of such heat-traces.

\subsection{Background field method quantisation}

The background field method (BFM) is a technique for calculations in quantum field theory based upon expanding fields about a fixed, arbitrary background that is used for quantising gauge field theories whithout losing explicit gauge invariance. Besides the advantage of maintaining explicit gauge invariance it greatly simplifies some computations. The BFM was introduced by DeWitt in 1967 \cite{dewitt1967bs} -- see \cite{abbott1981introduction,ABBOTT1983372} for historical and technical reviews. The main idea behind the BFM is to decompose a gauge field into a background field, $\bar{A}_\mu(x)$, and a quantum field, $a_\mu(x)$, which becomes the variable of integration in the functional integral. A gauge is fixed for the quantum field, yet despite this choice gauge invariance with respect to the background field is explicitly maintained. This method is quite general, yet of all quantum gauge choices the Feynman gauge plays a special role in the worldline formalism we shall employ below, allowing unification of the spin $1$ loop with spin $0$ and $1/2$ cases in the first quantised description \cite{Strassler:1992zr,Reuter:1996zm,ChrisRev} by decoupling external ghosts. This was essential for recent form factor decompositions of the off-shell gluon vertices which led to very compact results in comparison to the conventional methods based on the Ward identity analysis (see \cite{Ahmadiniaz:2012ie,Ahmadiniaz:2012xp,Ahmadiniaz:2016tsz,Ahmadiniaz:2016qwn} for the three- and four-gluon amplitudes).

With this method the computation of the the one-loop effective action (\ref{effact}) proceeds by expanding the gauge field about an arbitrary background so that $A_\mu(x)=\bar{A}_\mu(x)+a_\mu(x)$. With this shift the field strength tensor (\ref{Fstar}) is decomposed as  
\begin{equation}
F_{\mu\nu}=\bar{F}_{\mu\nu}+[\bar{D}_\mu,a_\nu]_{\star}-[\bar{D}_\nu,a_\mu]_{\star}-i[a_\mu,a_\nu]_{\star}
\end{equation}
where $\bar{F}_{\mu\nu}$ is the field strength tensor and $\bar{D}_\mu$ the covariant derivative computed from the background $\bar{A}_\mu$. With this shift the action becomes a functional of both $\bar{A}_\mu$ and $a_\mu$. As shown in \eqref{effact}, for the one-loop corrections to the effective action it is sufficient to study the piece quadratic in $a_\mu$ which takes the following form\footnote{Note that after cyclicity under $\int dx\,{\rm tr}\ldots$, the commutator $[D_\mu,\cdot]$ is anti-Hermitian.}
\begin{align}
    S_{\rm gauge}^{(2)}[\bar A,a]&=\frac{1}{g^2}\int_{\mathbb{R}^D}dx\ {\rm tr}\left(
    -a_\mu[\bar D_\nu,[\bar D_\nu,a_\mu]]-[\bar D_\mu,a_\mu]^2+2i\,a_\mu[\bar F_{\mu\nu},a_\nu]
    \right)\,,
 \label{s-2}
\end{align}
where multiplication is with the Moyal product throughout. The action $S^{(2)}_{\rm gauge}[\bar{A}, a]$ has invariance under the following (independent) transformations
\begin{eqnarray}
&&\delta_\alpha\bar{A}_\mu=0~,~~~~~~~~~~~~~~~~~~~~\,\delta_\alpha a_\mu=[D_\mu,\alpha]=[\bar{D}_\mu,\alpha]-i[a_\mu,\alpha]\label{transf-1}\\
&&\bar\delta_\alpha \bar{A}_\mu=[\bar D_\mu,\alpha]~,~~~~~~~~~~~~~\bar\delta_\alpha a_\mu=-i[a_\mu,\alpha]
\label{transf-2}
\end{eqnarray}
i.e.\ respectively quantum gauge invariance and background gauge invariance. 
We take advantage of this to fix the quantum gauge invariance (retaining gauge invariance with respect to the background field) by choosing the (Feynman-'t Hooft) gauge $[\bar{D}_\mu,a_\mu] = 0$, and we obtain the Faddeev-Popov determinant ${\rm Det} ([-\bar{D},[D,\cdot]])_{\star}$. Then, disregarding the linear term in $a_\mu$,  the appropriate contribution of the Faddeev-Popov ghosts to the total action at quadratic order reads,
\begin{align}
	S^{(2)}_{\rm ghost}[\bar{A},\bar{\lambda}, \lambda]=-\bar \lambda[\bar D_\mu,[\bar D_\mu,\lambda]]
\end{align}
where $\bar \lambda(x)$ and $\lambda(x)$ are one-component Grassmann fields valued in the fundamental representation of $u(N)$. Hence, (\ref{s-2}) is modified to the following compact form (where again all multiplication of fields is carried out with the Moyal $\star$-product)
\begin{align}
  \hspace{-2.5em}  S^{(2)}[\bar A,a, \bar{\lambda}, \lambda]&=\frac{1}{g^2}\int_{\mathbb{R}^D}dx\ {\rm tr}\left(
    -a_\mu[\bar D_\nu,[\bar D_\nu,a_\mu]]+2i\,a_\mu[\bar F_{\mu\nu},a_\nu]
    -\bar \lambda\bar D_\mu,[\bar D_\mu,\lambda]]\right).
\end{align}
As indicated by \eqref{effact}, functional integration over the quantum fields gives the one-loop effective action in terms of the determinants of the operators acting upon $a_\mu$ and $\lambda$ (we omit now the bars in the background field $\bar A_\mu$ and in the covariant derivative thereof)
\begin{align}
\Gamma[A]=\frac12\log {\rm Det}\Big\{\delta ^{2}S_{\rm gauge}\Big\}-\log{\rm Det}\Big\{\delta ^{2}S_{\rm ghost}\Big\}\,,
\end{align}
where
\begin{align}
	\delta ^{2}S_{\rm ghost}&=
	\frac{\delta^{2} S^{(2)}}{\delta  \lambda \, \delta  \bar{\lambda}}=
	-{\mathbb 1}_{\rm gauge}\otimes [D_\mu,[D_\mu,\cdot]]\,,\label{Dets}
	\\[2mm]
	\delta ^{2}S_{\rm gauge}&=
	\frac{\delta ^{2}S^{(2)}}{\delta a_{\mu}\delta a_{\nu}}=
	\left(-\delta_{\mu\nu}[D_\sigma,[D_\sigma,\cdot]]+2i\,[F_{\mu\nu},\cdot]\right)\otimes {\mathbb 1}_{\rm ghost}\,.\label{Detsgauge}
\end{align}
In both operators, the first factor acts on the space $\mathbb{R}^D\times L_2(\mathbb{R^D})$ of gauge fluctuations and the second, on $\mathbb{C}\times L_2(\mathbb{R^D})$, the space of ghost configurations. The fields $A_\mu(x),F_{\mu\nu}(x)$ in these operators take values in $u(N)$.

Using an integral representation of the determinants the effective action can be related to heat-traces after introducing IR ($m$) and UV ($\Lambda$) regulators to prevent the integral to diverge at small and large values of the Schwinger proper time $T$ \cite{Schwinger} 
\begin{align}
\Gamma[A]=-\frac{1}{2}\int_{\Lambda^2}^\infty \frac{dT}{ T}\,\e^{- T m^2}\Big({\rm Tr}\, \e^{- T \delta ^{2}S_{\rm gauge}}-2{\rm Tr}\, \e^{- T \delta ^{2}S_{\rm ghost}}\Big)
\label{efa}
\end{align}
In the next subsection we present a worldline determination of the heat-traces and test our construction by computing the one-loop $\beta$-function.  For the sake of such test we find it convenient to use a hard cut-off UV regularisation ($\Lambda$). However, one may as well use other UV regularisation schemes such as Dimensional Regularisation, which fits quite nicely in the worldline approach -- it dimensionally extends the Gamma functions that we shall see will appear from the Schwinger proper time integrals.

\subsection{Heat-trace representation in the worldline formalism}
We wish to give a first quantised representation of the heat-traces. We do this for the ghost sector in detail, since the steps are the same for the gauge fields (see the result (\ref{eq:YM-eff}) for this sector). The heat-trace for the ghost contribution to the action, which appears inside~\eqref{efa},
\begin{align}\label{trtr}
{\rm Tr}\, \e^{- T \delta^2 S_{\rm ghost}}=\int_{\mathbb{R}^D} dx
\,{\rm tr}\ \langle x\vert e^{- T \delta^2 S_{\rm ghost}}\vert x\rangle
\end{align}
is given by a transition amplitude $\langle x\vert \e^{- T \delta^2 S_{\rm ghost}}\vert x\rangle$ which can be computed in terms of a phase-space $\{x(t),p(t)\}$ path integral of a particle on the circle ~\citep{Bonezzi:2012vr, NCU1} with Hamiltonian $\delta^2 S_{\rm ghost}$. The symbol ${\rm tr}$ in \eqref{trtr} denotes the trace over the color indices in $\delta^{2}S_{\textrm{ghost}}$. The free kinetic operator of this particle has a zero mode on the circle which must be factored out. Different ways of factoring it out correspond to different particle propagators (Green functions), which however are known to lead to the same effective action in flat phase-space path integrals -- see~\cite{NCU1} for a detailed discussion of this issue for the worldline models used for non-commutative QFT's\footnote{In curved spaces, i.e. when the path integrals are represented by non-linear sigma models, this is a non-trivial issue that was discussed in~\cite{Bastianelli:2003bg, Corradini:2018lov}.}.  In the present work we adopt so-called string-inspired Green function, which corresponds to the zero mode being the center of mass of the path, $x_0$. We thus shift the worldline trajectories as $x(t)\to{x_0} + x(t)$, where now the quantum fluctuation satisfies $\int_0^ T dt\, x(t) = 0$.  With this bag of tricks we can rewrite the ghost contribution to the effective action as  
\begin{align}
{\rm Tr}\, \e^{- T \delta^2 S_{\rm ghost}}={\rm tr}\int_{\mathbb{R}^D} dx_0\,\int_{\rm PBC} \mathscr{D}x(t)\mathscr{D}p(t)\ \mathscr{P}\,\e^{-\int_0^ T dt\{-ip(t)\dot{x}(t)+H_{\rm ghost}^W(x(t),p(t))\}}
\label{heatk}
\end{align}
where PBC stands for the periodic boundary conditions for the trajectory $x(t)$, whereas no restrictions are imposed on $p(t)$. $H_{\rm ghost}^W$ is a $u_\star(N)$-valued function of the phase-space trajectories which thus requires the introduction of the path-ordering operator $\mathscr P$. This function $H_{\rm ghost}^W$ is obtained   after replacing $x\rightarrow x(t)$ and $\partial\rightarrow ip(t)$ in the Weyl-ordered expression of the Hamiltonian $\delta^2 S_{\rm ghost}$ (Weyl ordering is needed by the mid-point prescription of the particle path integral~\cite{Bastianelli:2006rx}). However, upon a formal Taylor expansion, one can easily show that the operator $e^{- T \delta^2 S_{\rm ghost}}$ is a combination of  the following mixed products (for some fields $\phi,\psi$)
\begin{align}
&\phi(x+i\theta\partial)\cdot\psi(x-i\theta\partial)\,,\nonumber\\
&\partial\cdot\phi(x+i\theta\partial)+\phi(x+i\theta\partial)\cdot\partial\,,\nonumber\\
&\partial\cdot\psi(x-i\theta\partial)+\psi(x-i\theta\partial)\cdot\partial\,,\nonumber
\end{align} 
all of which are already Weyl-ordered. The terms  $x+i\theta\partial$ and $x-i\theta\partial$ respectively correspond to left- and right-Moyal multiplications, i.e.
\begin{align}
(\phi \star \psi) (x) = \phi(x+i\theta \partial) \psi (x)= \psi(x-i\theta \partial) \phi (x)\,.
\end{align}
Thus, the full operator is Weyl-ordered, and no counterterms are needed in order to write it as a symmetrised expression of coordinates and momenta.

To write down the function $H_{\rm ghost}^W$ let us analyse the covariant derivative on the ghost field,
\begin{align}
[D_\mu,\lambda^a T^a]=\left(\partial_\mu \lambda^a
+\frac12\,f^{abc}\,\{A_\mu^b,\lambda^c\}-\frac{i}2\,d^{abc}\,[A_\mu^b,\lambda^c]\right)T^a
\,,
\end{align}
or, equivalently, as it acts on the adjoint fields $\lambda^{a}$,
\begin{align}
D^{\rm adj}_\mu\,\lambda^a&=\left(\delta^{ab}\,\partial_\mu
-\frac12\,f^{abc}\,[A_\mu^c(x-\theta p)+A_\mu^c(x+\theta p)]+\mbox{}\right.\nonumber\\
&\left.\mbox{}-\frac{i}2\,d^{abc}\,[A_\mu^c(x-\theta p)-A_\mu^c(x+\theta p)]\right)\lambda^b
\,.
\end{align}
The classical Hamiltonian counterpart of the operator $-(D^{\rm adj})^{2}$ then reads
\begin{align}
\left(H_{\rm ghost}^W\right)^{ab}=&\left\{\delta^{ad}\,p_\mu
+\frac12(i\,f^{adc}-d^{adc})\,A_\mu^c(x-\theta p)
+\frac12(i\,f^{adc}+d^{adc})\,A_\mu^c(x+\theta p)\right\}\nonumber\\
\times & \left\{\delta^{db}\,p^\mu
+\frac12(i\,f^{dbe}-d^{dbe})\,A^\mu{}^e(x-\theta p)
+\frac12(i\,f^{dbe}+d^{dbe})\,A^\mu{}^e(x+\theta p)\right\}\,.
\label{eq:Hgh}
\end{align}
With these results and redefining $x(t)\to\sqrt{ T} x(t)$ and $p(t)\to p(t)/\sqrt{T}$ for power counting convenience, with rescaled worldline time $t\in [0,1]$, the ghost contribution to the effective action can finally be written as  
\begin{align}
\Gamma_{\rm ghost} [A]&=-\frac{1}{2}\int_0^\infty \frac{d T}{ T}\ {\rm tr}\,\int_{\mathbb{R}^D}dx_0 \int \mathscr{D}x(t)\mathscr{D}p(t)
\ \mathscr{P}\,\e^{-\int_0^1 dt\  \left\{-ip\dot x+p^2+ T\, V(x,p)\right\}}~,
\label{eq:gh-eff}
\end{align}
where $\mathscr{P}$ represents the path-ordering, the trace is with respect to the color indices, and the potential $V(x,p)$ is the matrix obtained from the interacting part of the classical Hamiltonian~\eqref{eq:Hgh} with the scaled variables $x^{\mu} = x^{\mu}_{0} + \sqrt{T}x^{\mu}(t)$ and $p^{\mu} = p^{\mu}(t)/\sqrt{T}$. For simplicity, we have momentarily omitted the UV- and IR-regulators.

The $u_\star(N)$-valued potential $V$ can be split into the following five parts 
\begin{align}
V= V_+ + V_- + V_{++} + V_{--} +V_{\cal M}
\end{align}
where the subscripts indicate the number of left- and right-acting gauge fields $A$, whereas the fifth term is a mixed contribution which, as explained above, is linked to non-planar diagrams. Explicitly,
\begin{align}
V_+^{ab} &=  T^{-\frac12}(i\,f^{abc}+d^{abc})\,A_\mu^c (x_+(t))\,p^\mu(t)\\
V_-^{ab} &=  T^{-\frac12}(i\,f^{abc}-d^{abc})\,A_\mu^c (x_-(t))\,p^\mu(t)\\
V_{++}^{ab} &= \frac14(i\,f^{adc}+d^{adc})\,A_\mu^c (x_+(t))  (i\,f^{dbe}+d^{dbe})\,A^\mu{}^e (x_+(t)) \\
V_{--}^{ab} &= \frac14(i\,f^{adc}-d^{adc})\,A_\mu^c (x_-(t))  (i\,f^{dbe}-d^{dbe})\,A^\mu{}^e (x_-(t)) \\
V_{\cal M}^{ab} &= \frac14(i\,f^{adc}-d^{adc})\,A_\mu^c (x_-(t))  (i\,f^{dbe}+d^{dbe})\,A^\mu{}^e (x_+(t))
\nonumber\\ &+ \frac14(i\,f^{adc}+d^{adc})\,A_\mu^c (x_+(t))  (i\,f^{dbe}-d^{dbe})\,A^\mu{}^e (x_-(t))~,
\end{align}
where we have introduced the notation $x^\mu_\pm(t) = x_0^\mu +\sqrt{ T} x^\mu(t) \pm \theta^{\mu\nu} p_\nu (t)/\sqrt{ T}$.

The one-loop effective action can thus be used as a generating functional of effective vertices. In order to do that, one treats the potential as a perturbation of the free phase-space action, normalised as
\begin{align}
\int \mathscr{D}x(t)\mathscr{D}p(t) \,\e^{-\int_0^1 dt\  \left\{-ip(t)\dot x(t)+p^2(t)\right\}} =\frac{1}{(4\pi  T)^\frac D2}
\end{align}
so that
\begin{align}
\Gamma_{\rm ghost} [A]=-\frac{1}{2}\frac{1}{(4\pi)^{\frac{D}2}}\int_0^\infty \frac{d T}{ T^{1+\frac{D}2}}
\ {\rm tr}\,\int_{\mathbb{R}^D}dx\ \mathscr P
\ \left\langle \,\e^{- T\int_0^1 dt\ V(x(t),p(t))}\right\rangle\,.
\label{GamGhostMean}
\end{align}
The mean value is computed perturbatively by means of the following correlators (otherwise called worldline Green functions) derived in Appendix \ref{pbc}:
\begin{align}
\langle p_\mu(t) p_\nu(t')\rangle&=\frac12\,\delta_{\mu\nu}\,,\label{pp}\\
\langle x_\mu(t) x_\nu(t')\rangle&=2\,\delta_{\mu\nu}\,G(t-t')\,,\label{xx}\\
\langle p_\mu(t) x_\nu(t')\rangle&=i\,\delta_{\mu\nu}\,\dot G(t-t')\,,\label{px}
\end{align}
where $G(t)=-\frac12\left(|t| - t^{2}\right)$. 

As regards the gauge contribution to the one-loop effective action, we may similarly write
\begin{align}
\Gamma_{\rm gauge} [A]=-\frac{1}{2}\frac{1}{(4\pi)^{\frac{D}2}}\int_0^\infty \frac{d T}{ T^{1+\frac{D}2}}
\ {\rm tr}\,\int_{\mathbb{R}^D}dx\ \mathscr P
\ \left\langle \,\e^{- T\int_0^1 dt\
	\left\{\mathcal V(x(t),p(t))-2i\mathcal F(x(t),p(t))\right\}}\right\rangle\,,
\label{eq:YM-eff}
\end{align}
where the interactions include now an extra matrix structure due to Lorentz indices, so the trace refers to color as well as to space-time indices,
\begin{align}
\mathcal V^{ab}_{\mu\nu}&=\delta_{\mu\nu}\,V^{ab}(x(t),p(t))\,,\\
\mathcal{F}_{\mu\nu}^{ab}&=\frac{1}{2}\Big[(if^{abc}-d^{abc})F^c_{\mu\nu}(x-\theta p)+(if^{abc}+d^{abc})F^c_{\mu\nu}(x+\theta p)\Big]\,.
\label{eq:calF}
\end{align}
Note that the potential $\mathcal V$ is diagonal in the Lorentz indices and is essentially given by the same potential $V$ as for the ghost's contribution. On the other hand, the extra interaction term $\mathcal F$ is a non-trivial Lorentz matrix that is due to the field strength term in \eqref{Detsgauge}. Before leaving this section, it is worth comparing these Hamiltonians to the results of \cite{NCU1}. Due to the non-standard normalisation of the Lie algebra generator in the Abelian case it is necessary to use $f_{ijk} = 0$ and $d_{000} = 2$ to reproduce the worldline action used in the $U(1)$ case.

The gauge and ghost contributions to the  one- and two-point functions will be discussed in Section~\ref{secBeta}, where we will show that tadpole diagrams vanish and that the propagator contains, as expected, a logarithmically-divergent transverse contribution. Before doing that, in the next section we introduce a set of auxiliary fields for the worldline path integral, which will allow us to compactly write both ghost and Yang-Mills contributions to the full effective action as special cases of a single master formula and will help simplify the expansion of the effective action to second order in the gauge field.

\section{Auxiliary worldline fields}
\label{AWF}
Having arrived at a path integral representation of the one-loop effective action~(\ref{eq:gh-eff}, \ref{eq:YM-eff}) in phase-space, it is now possible to begin calculating the correlation functions. 
It is worth reiterating that in principle this can be achieved by making a straightforward expansion of the path ordered exponential of the action, followed by computation of the path integral over the worldline coordinates $x(t)$ and $p(t)$.

However, the worldline path integral as it stands is not yet optimal, since the potential entering into the action is matrix valued with respect to the gauge group indices and Lorentz indices and consequently requires the path ordering prescription for the exponentiated line integral. However, as  has now been applied in a number of cases \cite{Col1, Col2, JO1, JO2, ColTree, CreteProc}, building upon worldline representations of higher spin fields \cite{HSWL1, HSWL2, AsymT, ASWL1, ASWL2, Barduc1, HSDS1, Corradini:2010ia, HSDS2, HSC1, HSK1, HSK2} and differential forms \cite{Forms1, Forms2}, such an awkward approach can be avoided by the introduction of auxiliary worldline fields. These additional fields play the r\^{o}le of filling out the Hilbert spaces associated to extra degrees of freedom which in our case will represent gauge group information (termed ``colour fields'') and space-time forms (to be referred to as ``spin fields'' for reasons soon to become apparent). As such, we now introduce two independent sets of auxiliary worldline fields. 

\subsection{Auxiliary Colour fields}
The first, worldline colour fields, are exactly analagous to those used in \cite{Col1, Col2, JO1, JO2} and will absorb the gauge group indices, so we define a set of $N^{2}$ complex Grassmann variables\footnote{We could equally use bosonic fields with minimal changes to the steps that follow \cite{JO1}.}, which following \cite{ColTree, CreteProc} we denote by $\{\bar{c}^{a}\}_{a = 1}^{N^{2}}$ and $\{c_{a}\}_{a=1}^{N^{2}}$ that transform in the adjoint representation of $u(N)$. Choosing canonical Poisson brackets in Minkowski space
\begin{equation}
	\{\bar{c}^{a}, c_{b}\} = -i\delta^{a}_{b}
\end{equation}
facilitates the construction of a classical representation of the Lie algebra: define $S^{a} \equiv \bar{c}^{r}(T^{a})_{r}{}^{s}c_{s}$ and note
\begin{equation}
	\{S^{a}, S^{b}\}_{PB} = f^{abc}S^{c}.
\end{equation}
Upon quantisation -- where $\bar{c}^{a}$ and $c_{a}$ are replaced by operators acting, for example, on a coherent state basis with anti-commutation relations $\{\hat{\bar{c}}^{a},\hat{c}_{b}\} = \delta^{a}_{b}$ -- we then find
\begin{equation}
	[\hat{S}^{a}, \hat{S}^{b}] = if^{abc}\hat{S}^{c}.
	\label{LieClass}
\end{equation}
The Minkowski space action compatible with these brackets is
\begin{equation}
	S[\bar{c}, c] = i\int_{0}^{1} \bar{c}^{a}\dot{c}_{a} \, d\tau.
\end{equation}
We observe that the Green function of these worldline fields, derived from the first order kinetic term, is essentially the step function. This is sufficient to generate a path ordering prescription, so that it need no longer be imposed by hand, whilst the anti-periodic boundary conditions implement the trace over colour indices. Moreover, given (\ref{LieClass}), the colour fields can be used to absorb the gauge group indices of the matrix valued potential in the action. 

This leads to a modification of the worldline action which on the unit circle becomes (the Lorentz space identity matrix has been suppressed to avoid cluttering)
\begin{align}
	S[p, x, \bar{c}, c, a] = \int_{0}^{1}dt\bigg[&-i p \cdot \dot{x} + p^{2}+ \bar{c}^{a}\dot{c}_{a} +T V_{a}{}^{b}\left(\bar{c}^{a} c_{b} + \frac{1}{2}\delta^{a}_{b}\right)   \nonumber \\
	&- 2iT (\mathcal{F}_{\mu \nu})_{a}{}^{b}\left(\bar{c}^{a} c_{b} + \frac{1}{2}\delta^{a}_{b}\right) + ia(t)\left(\bar{c}^{a}c_{a} -  n\right) \bigg],
	\label{Scgauged}
\end{align}
where $(\mathcal{F}_{\mu \nu})^{ab}$ is given in~\eqref{eq:calF}. 
Note that we have Weyl ordered the terms involving the colour fields (which produces the factors $\frac{1}{2}\delta_{a}^{b}$) which is part of the path integral regularisation; such factors were absent in previous applications of worldline colour fields based on $SU(N)$ or $SO(N)$ symmetry groups due to tracelessness of the matrices that appeared in the action. 

In the action (\ref{Scgauged}) we have also gauged the $U(1)$ phase symmetry 
by introducing a (worldline) gauge field $a(t)$. We will immediately fix this gauge symmetry by choosing $a(t) = \vartheta$, a constant modulus distinguishing gauge inequivalent configurations. Examination of large gauge transformations shows that $\vartheta \in [0, 2\pi]$ that must be integrated over with convenient normalisation of the measure $\frac{d\vartheta}{2\pi}$. The constant $n \equiv \varrho - \frac{N^{2}}{2}$ projects onto an irreducible representation of the colour Hilbert space ($\varrho = 1$ will project onto the adjoint representation). 
 To see this, note that the auxiliary fields $c_a$
 correspond to a set of $N^{2}$ fermionic oscillators that generate a $2^{N^{2}}$-dimensional Hilbert space
\begin{align}
{\cal H}_{C} =\bigoplus_{p=0}^{N^{2}} {\cal H}_\varrho\,,\quad {\rm dim}({\cal H}_\varrho) ={\rm tr}\, ({\mathbb 1}_\varrho)= {{N^{2}}\choose{}\varrho}
\label{eq:H}
\end{align}
where ${\cal H}_\varrho$ denotes the subspace with $c$-occupation number $\varrho$: in a coherent state basis we realise the anti-commutation relations by the identification $\hat{\bar{c}}^{a} \rightarrow \bar{c}^{a}$, $\hat{c}_{a} = \frac{\partial}{\partial \bar{c}^{a}}$ for Grassmann variables $\bar{c}^{a}$ and then an arbitrary state in such a Hilbert space is represented by a wavefunction with completely anti-symmetric components $\varphi_{a_{1}\ldots a_{p}}(x)$,
\begin{equation}
	\phi(x, \bar{c}) = \varphi(x) + \varphi_{a_{1}}(x)\bar{c}^{a_{1}} + \cdots + \varphi_{a_{1}\ldots a_{p}}(x)\bar{c}^{a_{1}}\cdots\bar{c}^{a_{p}} + \ldots \varphi_{a_{1}\ldots a_{N^{2}}}(x)\bar{c}^{a_{1}}\cdots\bar{c}^{a_{N^{2}}}.
	\label{PhiCol}
\end{equation}
The number operator is given by $\hat N = {\hat{\bar{c}}^{a}} \hat{c}_{a}=\bar c^a\frac{\partial}{\partial \bar c^a}$ which allows projection operators to be written
\begin{align}
{\mathbb P}_{\varrho} = \int_0^{2\pi} \frac{d\vartheta}{2\pi}\, e^{i\vartheta (\varrho-\hat N)} \ \Longrightarrow\ {\rm tr}_{\cal H} \big( {\mathbb P}_\varrho \big)  = {\rm tr}\, ({\mathbb 1}_\varrho)~.
\label{eq:Pr}
\end{align}
This selects from (\ref{PhiCol}) the wavefunction transforming in the representation with $\varrho$ anti-symmetric indices. The path integral representation of this projector is implemented in (\ref{Scgauged}) where it arises from the gauge fixing of the local $U(1)$ symmetry gauged by the worldline fields $a(\tau)$ discussed above. This provides 
\begin{equation}
	\int \mathscr{D}p \oint \mathscr{D}x \oint \mathscr{D}[\bar{c},c] \int_0^{2\pi} \frac{d\vartheta}{2\pi} \, \tr \mathscr{P} \exp{\left(-S[p, x, \bar{c}, c, \vartheta]\right)}
\end{equation} 
with anti-periodic boundary conditions on the Grassmann colour fields and periodic boundary conditions on the particle trajectories. Note, however, that we still require a path ordering prescription for the Lorentz indices that appear inside the worldline action. We discuss how this can be circumvented in a similar way to the gauge group indices in the following section.

\subsection{Auxiliary Spin fields}
It would be advantageous to remove the path ordering associated with the Lorentz matrix structure of the worldline action by also absorbing their indices by the introduction of auxiliary fields. A preliminary version of this approach was presented in early work calculating the one-loop effective action in commutative space \cite{Sp1, Pro1} building upon the original proposal of \cite{Strassler:1992zr}, that was sufficient to generalise the well-known ``cycle replacement rules'' relating amplitudes in scalar and spinor quantum electrodynamics to the case of a spin-one particle in a virtual loop -- for a concise description of this approach see \cite{ChrisRev, OurRep}.  However, in the above-mentioned works the projection onto the correct subspace of the enlarged Hilbert space required to include the spin degrees of freedom was less elegant (involving the introduction of an additional mass-term and parity operator) than the updated technique we will use here.  This projection is essentially the same as that used to pick out a wavefunction transforming in an irreducible representation of the gauge group presented in the previous section. 

Recall that the worldline path integrals reproduce the functional determinants of the differential operators given in equations~\eqref{Dets} and \eqref{Detsgauge}, which can be re-written as 
\begin{equation}
		 \left(-{\mathbb{1}} [D_\sigma,[D_\sigma,\cdot]]+2i \zeta \,[{\mathbb F},\cdot]\right)_{\star}
		 \label{opsZeta}
\end{equation}
with $\zeta = 1$ for the gauge degrees of freedom (where ${\mathbb 1}$ is the space-time identity) and $\zeta = 0$ for the ghosts (where ${\mathbb 1} =1$), and where ${\mathbb F}$ represents the field strength. 
Following the treatment of the above section, we introduce additional Grassmann operators with space-time indices, $\hat{\bar{\psi}}^{\mu}$ and $\hat{\psi}^{\mu}$ with anti-commutator $\{\hat{\bar{\psi}}^{\mu}, \hat{\psi}_{\nu}\} = \delta^{\mu}{}_{\nu}$. These operators generate a $2^{D}$-dimensional Hilbert space spanned by wavefunctions with an expansion in terms of Lie-algebra  valued $r$-forms, $\phi_{\mu_{1}\ldots \mu_{r}}$ for $r \in \{0,\ldots, D\}$,
\begin{equation}
	\Phi(x,\bar{\psi}) = \phi(x) + \phi_{\mu_{1}}(x)\bar{\psi}^{\mu_{1}} + \cdots + \phi_{\mu_{1}\ldots \mu_{r}}(x )\bar{\psi}^{\mu_{1}}\cdots\bar{\psi}^{\mu_{r}} + \ldots \phi_{\mu_{1}\ldots \mu_{D}}(x)\bar{\psi}^{\mu_{1}}\cdots\bar{\psi}^{\mu_{D}}.
	\label{eq:graded-wf}
\end{equation}
each of which gives a subspace of dimension ${\scriptscriptstyle \nCr{D}{r}}$. The projection operator that selects the component with $r$ anti-symmetric indices is built from the number operator $\hat{L} = \hat{\bar{\psi}}^{\mu}\hat{\psi}_{\mu}$ as
\begin{align}
{\mathbb P}_r = \int_0^{2\pi} \frac{d\varphi}{2\pi}\, e^{i\varphi (r-\hat L)} \ \Longrightarrow\ {\rm tr}_{\cal H} \big( {\mathbb P}_r \big)  = {\rm tr}\, ({\mathbb 1}_r)~.
\label{eq:Pr2}
\end{align}
With these operators we define a Hamiltonian 
\begin{align}
\hat{H} := -\mathbb 1 [D_\sigma,[D_\sigma,\cdot]] + 2i \, \hat{\bar \psi}^\mu [F_{\mu\nu}, \cdot] \hat{\psi}^\nu. 
\end{align}
which acts on the graded wave function~\eqref{eq:graded-wf}. It acts as~\eqref{Dets} ($\zeta = 0$) on the subspace of zero-forms, where ghost fields live, and as~\eqref{Detsgauge} ($\zeta = 1$) on the space of one-forms, where gauge fields live. 
In a coherent state basis we can  identify $\hat{\bar{\psi}} \rightarrow \bar \psi$ and $\hat{{\psi}} \rightarrow \frac{\partial}{\hspace{-0.2em}\raisebox{-0.08em}{ \scriptsize $\partial \bar{\psi}$ } \hspace{-0.2em}}$, so that
\begin{align}
	\hat{H}\phi_{\mu}(x)\bar{\psi}^{\mu} &= \left(-\delta_{\mu\nu} [D_\sigma,[D_\sigma,\phi_{\nu}(x)]] + 2i [F_{\mu\nu},\phi_{\nu}(x)]\right)\bar\psi^{\mu} \\
	\hat{H}\phi(x) &= - [D_\sigma,[D_\sigma,\phi(x)]]~.
\end{align}
The projection operators can be implemented at the level of the path integral in order to select the appropriate subspace for the gauge and ghost degrees of freedom. We introduce Grassmann functions $\bar{\psi}^{\mu}(\tau)$ and $\psi^{\mu}(\tau)$ in the coherent state basis with kinetic term $\int_{0}^{1}d\tau\, \bar{\psi}^{\mu}\dot{\psi}_{\mu}$ which yields a Poisson bracket compatible with the quantum anti-commutation relations\footnote{This can furnish us with a classical representation of the Lorentz algebra: given generators $(M^{\mu\nu})_{\alpha \beta}$ with Lorentz indices $\alpha$, $\beta$ we may define $\mathcal{M}^{\mu\nu}:= \bar{\psi}^{\alpha}(M^{\mu\nu})_{\alpha \beta}\psi^{\beta}$ and note their Poisson bracket is given by $\{\mathcal{M}^{\mu\nu}, \mathcal{M}^{\rho \sigma}\}_{PB} = i\left(g^{\mu\rho}\mathcal{M}^{\nu\sigma} - g^{\nu\rho}\mathcal{M}^{\mu\sigma} - g^{\mu\sigma}\mathcal{M}^{\nu\rho} + g^{\nu\sigma}\mathcal{M}^{\mu\rho}\right)$.}. The Green function of these fields is again sufficient to generate the path ordering and anti-periodic boundary conditions produce the trace over Lorentz indices. Moreover, the anti-symmetry of the field strength indices automatically yields Weyl ordering. In the path integral described below, the projection~\eqref{eq:Pr2} can be realized in the same way as for the colour degrees of freedom, i.e. by introducing a $U(1)$ gauge field $\phi(t)$ which in turn can be gauge-fixed to the angle variable $\varphi$. 

Grassmann fields as well as bosonic gauge fields are routinely applied in worldline representations of fields of arbitrary spin~\cite{AsymT, Barduc1, Bastianelli:2002qw, ASWL1, ASWL2,  HSWL1, HSWL2, HSDS1}. In such cases the resulting free particle action usually enjoys a worldline supersymmetry between the bosonic coordinates $x^{\mu}$ and the fermionic fields $\psi^{\mu}$ -- that can be extended to incorporate the colour fields -- which has profound consequences \cite{ChrisRev, CreteProc}; although in general this supersymmetry can survive the transition to non-commutative space, it turns out not to apply in the current case (see \eqref{Scpsigauged} below) where the supersymmetry algebra does not close once a coupling of the point particle to the gauge field is included (as would also be the case in commutative space). 

\subsection{The worldline representation of the effective action}
Putting together the ingredients introduced in the previous subsections we arrive at the final version of the worldline action to be used in our non-commutative setup, i.e.
\begin{align}
&	S_{\varrho, r}[p, x,  \bar{\psi}, \psi,\bar{c}, c, \varphi, \vartheta]= \int_{0}^{1}dt\bigg[-i p \cdot \dot{x} + p^{2} + \bar{c}^{a}\dot{c}_{a} + \bar{\psi} \cdot \dot{\psi} + T V_{a}{}^{b}\left(\bar{c}^{a} c_{b} + \frac{1}{2}\delta^{a}_{b}\right) \nonumber \\
	&- 2iT \bar{\psi}^{\mu}(\mathcal{F}_{\mu \nu})_{a}{}^{b}\psi^{\nu}\left(\bar{c}^{a} c_{b} + \frac{1}{2}\delta^{a}_{b}\right) + i\vartheta\left(\bar{c}^{a}c_{b} -  n\right) + i\varphi\left(\bar{\psi} \cdot \psi - s\right)\bigg]\,,
	\label{Scpsigauged}
\end{align}
where the Chern-Simons charge $s \equiv r - \frac{D}{2}$ fixes the degree of the form making up the wavefunction $\Phi(x, \bar{\psi})$. The factor $- \frac{D}{2}$ appears due to Weyl-ordering the number operator:
as mentioned above, for the gauge sector we set $r = 1$ and for the ghosts we fix $r = 0$. Finally $\varphi$ is the $U(1)$ modulus left over after gauge fixing that is to be integrated over; it is to $\phi(t)$ as $\vartheta$ is to $a(t)$.

Hence, denoting by $\Gamma[A]_{\varrho,r}$ the complete path integral including auxiliary fields that project onto the representation of $U(N)$ with $\varrho$ fully anti-symmetric indices and the Lorentz sector of the Hilbert space associated to $r$-forms,
\begin{align} 
	\hspace{-5em}\Gamma[A]_{\varrho,r} &:=-\frac{1}{2} \int_{0}^{\infty}\frac{d  T}{(4 \pi)^{\frac{D}{2}} T^{1 + \frac{D}{2}}} \int \mathscr{D}p \oint \mathscr{D}x \oint\mathscr{D}[\bar{\psi}, \psi] \oint \mathscr{D}[\bar{c}, c]  \int_{0}^{2\pi} \frac{d \varphi}{2\pi}\,  \int_0^{2\pi} \frac{d\vartheta}{2\pi} \nonumber\\&\times \exp{\left(-S_{\varrho, r}[p, x, \bar{\psi}, \psi, \bar{c}, c, \varphi, \vartheta]\right)}\,,
	\label{Gammapq}
\end{align} 
then the one-loop effective action has path integral representation
\begin{equation}
	\Gamma[A] = \Gamma[A]_{1, 1} - 2\Gamma[A]_{1,0}.
	\label{GammapqA}
\end{equation}
In the path integrals, all Grassmann variables have anti-periodic boundary conditions, the bosonic embedding coordinates have periodic boundary conditions and the momenta are unrestrained. Note that at last we have arrived at a path integral representation which contains a pure exponential function, since the auxiliary fields produce the required path ordering. This represents a major advantage of the auxiliary field formulation. Equation (\ref{GammapqA}) will form the basis of our calculation of the $\beta$-function by expanding the two contributions to quadratic order in the gauge field. 

There is one final modification to the path integrals defined above. It is convenient to absorb the $\vartheta$- and $\varphi$-dependence of the worldline action by redefining the worldline fields according to
\begin{equation}
	\begin{pmatrix} \bar{c}^{a}(t) \\ \bar{\psi}^{\mu}(t)\end{pmatrix} \rightarrow \begin{pmatrix} \bar{c}^{a}(t) \\ \bar{\psi}^{\mu}(t)\end{pmatrix} \begin{pmatrix}e^{-i\vartheta t} & 0 \\ 0 & e^{-i \varphi t}\end{pmatrix}; \qquad \begin{pmatrix} c^{a}(t) \\ \psi^{\mu}(t) \end{pmatrix} \rightarrow \begin{pmatrix}e^{i\vartheta t} & 0 \\ 0 & e^{i \varphi t}\end{pmatrix} \begin{pmatrix} c^{a}(t) \\ \psi^{\mu}(t)\end{pmatrix}
\end{equation}
which turns the original \textit{anti-periodic} boundary conditions on the Grassmann fields $\bar{\psi}^{\mu}$, $\psi^{\mu}$ and $\bar{c}^{a}$, $c_{a}$ into \textit{twisted} boundary conditions ($\TBC$)
\begin{equation}
\begin{split}
c^a(1) &=-e^{i\vartheta} c^a(0)\,,\quad \bar c^a(1) =-e^{-i\vartheta} \bar c^a(0)\\
\psi^\mu (1) &=-e^{i\varphi} \psi^\mu(0)\,,\quad \bar\psi^\mu (1) =-e^{-i\varphi} \bar\psi^\mu(0) ~.
\end{split}
\label{eq:BCs}
\end{equation}
 Putting this all together we get an equivalent representation of (\ref{Gammapq}) 
\begin{align}
\Gamma[A]_{\varrho, r} &= -\frac{1}{2} \int_{\Lambda^{-2}}^{\infty} \frac{dT \,e^{-m^{2}T}}{(4\pi)^{\frac{D}{2}}T^{\frac{D}{2} + 1}} \int d^{D}x\int_{0}^{2\pi} \frac{d \varphi}{2\pi} \int_{0}^{2\pi} \frac{d \vartheta}{2\pi} \, e^{i\vartheta \big(\varrho - \frac{N^{2}}{2}\big)}e^{i \varphi \big(r - \frac{D}{2}\big)}\nonumber\\ &\times \int \mathscr{D}p  \int_{\PBC}\hspace{-1em} \mathscr{D}x \int_{\TBC} \hspace{-1em} \mathscr{D}[\bar{\psi},\psi] \int_{\TBC}\hspace{-1em} \mathscr{D}[\bar{c}, c] \,\e^{-S[p, x, \bar{c}, c, \bar{\psi}, \psi]},
	\label{GammaTwist}
\end{align}
where the gauge fixed action with shifted worldline auxiliary fields is simplified to
\begin{align}
	S[p, x, \bar{c}, c, \bar{\psi}, \psi] =  \int_{0}^{1}dt\bigg[ & -i p \cdot \dot{x} + p^{2} + \bar{c}^{a}\dot{c}_{a} + \bar{\psi} \cdot \dot{\psi} + TV_{a}{}^{b}\left(\bar{c}^{a} c_{b} + \frac{1}{2}\delta^{a}_{b}\right) \nonumber\\& - 2i T \bar{\psi}^{\mu}(\mathcal{F}_{\mu \nu})_{a}{}^{b}\psi^{\nu}\left(\bar{c}^{a} c_{b} + \frac{1}{2}\delta^{a}_{b}\right) \bigg]~.
\label{Sfixshift}
\end{align}
The action (\ref{Sfixshift}) and path integral (\ref{GammaTwist}) represent one of the main results of this paper -- a first quantised representation of the effective action where all degrees of freedom are treated equally and the gauge and ghost sectors are unified without the need for a manual path ordering. In the next section we compute the path integral over the auxiliary fields, expanding the effective action up to quadratic order in the gauge field. 

\section{Planar divergences of the effective action}
The planar divergences that contribute to the $\beta$-function arise from terms involving products of gauge fields -- or the field strength tensor -- evaluated at either $(x + \theta \cdot p)$ or $(x - \theta \cdot p)$. Here we begin the calculation of the $\beta$-function based on the two point function, at quadratic order in the gauge field, so we expand the effective action (\ref{GammaTwist}) to second order for $p = 1$ and $r = 0$, $1$ and discount products of fields with respective arguments $(x \pm \theta \cdot p)$ and $(x \mp \theta \cdot p)$. In this section we carry out this expansion and compute the path integrals over the auxiliary worldline fields, leaving the eventual computation of the remaining path integral over phase-space and the proper time integral for Section \ref{secBeta}.

Referring back to (\ref{GamGhostMean}) and (\ref{eq:YM-eff}), and implementing the auxiliary fields as described above, we can write the more general path integral as an expectation value (with $\left<1\right> = 1$):
\begin{align}
\hspace{-1em}	 \Gamma[A]_{\varrho, r} =& -\frac{1}{2}\int_{\Lambda^{-2}}^{\infty} \frac{dT \,e^{-m^{2}T}}{(4\pi)^{\frac{D}{2}}T^{\frac{D}{2} + 1}} \int d^{D}x \int_{0}^{2\pi} \frac{d \varphi}{2\pi} \int_{0}^{2\pi} \frac{d \vartheta}{2\pi} \, e^{i\vartheta \big(\varrho - \frac{N^{2}}{2}\big)}e^{i \varphi \big(r - \frac{D}{2}\big)} \nonumber \\
	\hspace{-1em}	 &\times\left(2\cos\frac{\vartheta}{2}\right)^{N^{2}}\left(2 \cos\frac{\varphi}{2}\right)^{D}\left< e^{-T\int_{0}^{1} d\tau \left[ \left( V_{a}{}^{b}  -2i \bar{\psi}^{\mu}(\mathcal{F}_{\mu\nu})_{a}{}^{b}\psi^{\nu}\right)\left( \bar{c}^{a} c_{b} + \frac{1}{2}\delta_{a}^{b}\right)  \right]} \right>
	 \label{GamExp}
\end{align}
where the trigonometric expressions arise due to the normalisation of  the free path-integrals over the auxiliary fields
\begin{align}
	\int_{TBC} \mathscr{D}[\bar{\psi},\psi]\, e^{-\int_0^1 dt\, \bar{\psi}^{\mu} \dot \psi_{\mu}}  &= \Big(2\cos\frac{\varphi}{2}\Big)^D\,,\\
	\int_{\TBC}\mathscr{D}[\bar{c}, c]\, e^{-\int_{0}^{1}\, \bar{c}^{a}\dot{c}_{a} dt} &= \left(2\cos\frac{\vartheta}{2}\right)^{N^{2}}~.
\end{align} 
Above, along with the Green's functions defined in (\ref{pp})-(\ref{px}), we make use of the Green functions associated to the auxiliary fields that are read off from their kinetic actions. With the employed twisted boundary conditions, these are:
\begin{align}
\big\langle \psi_{\mu}(t) \bar{\psi}^{\nu}(s)\big\rangle_c &=\delta_\mu^\nu \Delta(t-s;\varphi)\,, \\
\left< {c}_{a}(t) \, \bar c^{b}(s)\right> &= \delta^{a}_{b}\Delta(t - s, \vartheta)\,,
\end{align}
where 
\begin{align}
&\Delta(t-s;\omega):= \frac{1}{2\cos\frac{\omega}{2}}\Bigl[ e^{i\frac{\omega}{2}} \Theta(t-s) - e^{-i\frac{\omega}{2}} \Theta(s-t)\Bigr]\,.
\label{eq:green-cc}
\end{align}
Note that in the limit $\omega \rightarrow 0$ we recover $\Delta(t-s ; 0) = \frac{1}{2}\sigma(t -s )$ that is appropriate for a kinetic term first order in derivatives with anti-periodic boundary conditions. We provide some further discussion of these worldline Green functions in the Appendix~\ref{pbc}. Below we consider the expansion of this expectation value up to quadratic order in the gauge field and will evaluate (\ref{GamExp}) for $\varrho = 1$ and $r = 1$, $0$ for the gauge and ghost contribution respectively. However, we can calculate these two contributions simultaneously by keeping $r$ arbitrary until the end of the calculation, that represents a significant advantage granted by the auxiliary worldline field approach.

\subsection{Single vertex insertion}
We first demonstrate that there are no planar contributions that are linear in the gauge field. We expand the exponential of (\ref{GamExp}) to first order, which introduces an insertion of single vertices built from $V$ and $\cal{F}$. Firstly note that the first-order contribution from the $\mathcal{F}$-term involves $\left<\bar{\psi}^{\mu}(t)\psi^{\nu}(t)\right> \mathcal{F}_{\mu\nu}$ which is proportional to $\delta^{\mu\nu}F_{\mu\nu}(x + \theta \cdot p) = 0$ due to tracelessness of the field strength tensor. In fact to build anything non-zero from this part of the action we need to go to at least quadratic order in $\mathcal{F}$. 

On the other hand the $V$-part of the action contributes at $\mathcal{O}(A)$ and $\mathcal{O}(A^{2})$. The two (left and right) linear contributions in $A$ are distinguished by $\pm$ and read
\begin{equation}
\begin{split}
&\frac{1}{2}\int_{\Lambda^{-2}}^{\infty} \frac{dT \, e^{-m^{2}T}}{(4\pi)^{\frac{D}{2}}T^{\frac{D+1}{2}}} \int d^{D}x  \, (i f_{abc} \pm d_{abc})\int_{0}^{2\pi} \frac{d \varphi}{2\pi} \int_{0}^{2\pi} \frac{d \vartheta}{2\pi} \, e^{i\vartheta \big(1 - \frac{N^{2}}{2}\big)}e^{i \varphi \big(r - \frac{D}{2}\big)}  
 \\
	&\times \left(2\cos\frac{\vartheta}{2}\right)^{N^{2}}\left(2 \cos\frac{\varphi}{2}\right)^{D} \int_{0}^{1}dt \left<A_{\mu}^{c}(x)\, p_{\mu}\right>\left< \bar{c}^{a}(t)c_{b}(t) + \frac{1}{2}\delta_{b}^{a}\right>~,
	\label{FOA}
\end{split}
\end{equation}
where since the operatorial dependence of the potential is shifted away by the integral over the zero mode we are free to evaluate it at the point $x$: this property obviously holds at any order in $p$. 

With $\left< \bar{c}^{a}(t)c_{b}(t) + \frac{1}{2}\delta_{a}^{b}\right> = \delta_{a}^{b}\left(-\frac{i}{2} \tan\frac{\vartheta}{2} + \frac{1}{2}\right)$ we immediately remove the contributions from the anti-symmetric structure constants (the remaining contribution would therefore vanish in a commutative $SU(N)$ theory). Furthermore, the integrals over the moduli read
\begin{align}
	&\int_{0}^{2\pi} \frac{d\varphi}{2\pi}\left(2 \cos\frac{\varphi}{2}\right)^{D}e^{i \varphi \big(r - \frac{D}{2}\big)} = \nCr{D}{r}; \quad  0 \leqslant r \leqslant D, \\
	\begin{split}
	&\int_{0}^{2\pi} \frac{d\vartheta}{2\pi}\left(2 \cos\frac{\vartheta}{2}\right)^{N^{2}}\left(\frac{i}{2} \tan \frac{\vartheta}{2}\right)^{n} e^{i \vartheta\left(1 - \frac{N^{2}}{2}\right)}\\
	 &= \frac{1}{2^{n}}\oint \frac{dz}{2\pi i}\left(1 + \frac{1}{z}\right)^{N^{2} - n}\left(1 - \frac{1}{z}\right)^{n}  
	 	 = \frac{N^{2} - 2n}{2^{n}};\quad 0 \leqslant n \leqslant N^{2}
	\end{split}
	 \label{Ints1}
\end{align}
so that the full integral over $\vartheta$ is equal  $1$, and (\ref{FOA}) thus becomes
\begin{equation}
	\frac12 \nCr{D}{r} \int_{\Lambda^{-2}}^{\infty} \frac{dT e^{-m^{2}T}}{(4\pi)^{\frac{D}{2}}T^{\frac{D+1}{2}}} (\pm)d_{aac} \, \int d^{D}x \int_{0}^{1}dt \left<A_{\mu}^{c}(x) p_{\mu}(t)\right>.
\end{equation}
We set $r = 0$ for the ghost sector and $r = 1$ for the gauge sector; the latter contribution generates an additional factor of $D$ that is implementing the trace over Lorentz indices (recall that the trace is generated by integrating over the auxiliary variables $\bar{\psi}$, $\psi$, which by this stage has become an integration over the modulus $\varphi$). 




\subsubsection{Terms quadratic in $A$}
At first order in expansion of the exponential there are also four terms quadratic in the gauge potential, two planar terms ($++$ and $--$) and two non planar ones ($+-$ and $-+$). The two planar contributions read
\begin{align}
	&\frac{1}{8} \int_{\Lambda^{-2}}^{\infty} \frac{dT \, e^{-m^{2}T}}{(4\pi)^{\frac{D}{2}}T^{\frac{D}{2}}}\, \int d^{D}x \Big(f_{acd}f_{bce}+d_{acd}d_{bce}\pm i(f_{acd}d_{bce}-f_{bcd}d_{ace})\Big)\nonumber\\
&\times\int_{0}^{2\pi} \frac{d \varphi}{2\pi} \int_{0}^{2\pi} \frac{d \vartheta}{2\pi}\,e^{i\vartheta \big(1 - \frac{N^{2}}{2}\big)}e^{i \varphi \big(r - \frac{D}{2}\big)} \left(2\cos\frac{\vartheta}{2}\right)^{N^{2}}\left(2 \cos\frac{\varphi}{2}\right)^{D}\nonumber\\ 
&\times\int_{0}^{1} dt\, \left<A_\mu^{d}(x)\,A_\mu^{e}(x)\right> \left< \bar{c}^{a}(t)c_{b}(t) + \frac{1}{2}\delta_{a}^{b}\right>\,,
\end{align}
where once again the operatorial part of the argument of the gauge fields gets shifted away.
The expectation value of the auxiliary fields yields a result proportional to $\delta^{a}_{b}$ which makes the latter two terms involving the structure constants vanish by symmetry: the two planar contributions are thus equal. Subsequently, integrating over the auxiliary fields using (\ref{Ints1}) provides
\begin{equation}
	\frac{1}{8}\nCr{D}{r}\int_{\Lambda^{-2}}^{\infty} \frac{dT e^{-m^{2}T}}{(4\pi)^{\frac{D}{2}}T^{\frac{D}{2}}}\left(f_{acd}f_{ace}+d_{acd}d_{ace}\right) \int d^{D}x \int_{0}^{1} dt \left<A_{\mu}^{d}(x)A_{\mu}^{e}(x)\right>.
\end{equation}
where we will again need either $r = 0$ ($r = 1$) for the ghost (gauge) contribution.

\subsection{Double vertex insertion}
At second order in the expansion of the exponential we find terms quadratic, cubic and quartic in the field; for the $\beta$-function it suffices to determine the quadratic pieces. To begin with, we get our first contribution from the field strength tensor involving the auxiliary spin fields. The planar terms are:
\begin{align}
	&\frac{1}{4}\int_{\Lambda^{-2}}^{\infty} \frac{dT e^{-m^{2}T}}{(4\pi)^{\frac{D}{2}}T^{\frac{D-2}{2}}} \int d^{D}x \left(\pm d_{abc} + if_{abc}\right)\left(\pm d_{rst} + if_{rst}\right)\nonumber\\
	&\times \int_{0}^{2\pi} \frac{d \varphi}{2\pi} \int_{0}^{2\pi} \frac{d \vartheta}{2\pi} \, e^{i\vartheta \big(1 - \frac{N^{2}}{2}\big)}e^{i \varphi \big(r - \frac{D}{2}\big)}\left(2\cos\frac{\vartheta}{2}\right)^{N^{2}}\left(2 \cos\frac{\varphi}{2}\right)^{D}\nonumber\\
	&\times \int_{0}^{1}dt \int_{0}^{1}dt^{\prime} \,\left<F_{\mu\nu}^{c}(x_\pm) F_{\alpha \beta}^{ t}(x'_\pm)\right>\left<\bar{\psi}^{\mu}(t)\psi^{\nu}(t)\bar\psi^{\alpha}(t^{\prime}){\psi}^{\beta}(t^{\prime})\right>\nonumber \\
	&  \hphantom{ \left(2\cos\frac{\vartheta}{2}\right)^{N^{2}}\left(2 \cos\frac{\varphi}{2}\right)^{D} \int_{0}^{1}dt \int_{0}^{1}dt^{\prime} }\times \left< \left(\bar{c}^{a}(t)c_{b}(t) + \frac{1}{2}\delta^{a}_{b} \right) \left(\bar{c}^{r}(t^{\prime})c_{s}(t^{\prime}) + \frac{1}{2}\delta^{r}_{s}\right)\right>~,\nonumber \\
\end{align}
where $x'_\pm :=x_\pm(t')$.
Now we need not contract $\bar{\psi}$ and $\psi$ at equal times due to tracelessness of the field strength tensors (which is why this term does not give anything at first order). This leaves us with the surviving contraction $\left<\bar{\psi}^{\mu}(t)\psi^{\nu}(t) \bar \psi^{\alpha}(t^{\prime}){\psi}^{\beta}(t^{\prime})\right> = -\delta^{\mu\beta}\delta^{\alpha\nu}\Delta(t - t'; \varphi)\Delta(t' - t; \varphi) = \delta^{\mu\beta}\delta^{\alpha\nu}\left(2 \cos \frac{\varphi}{2}\right)^{-2}$. The Kronecker $\delta$'s contract the indices on the field strength tensors to form $\tr\left[F^{c}(x_\pm) \cdot F^{t}(x'_\pm)\right]$. Turning to the colour fields we need
\begin{align}
	&\left< \left(\bar{c}^{a}(t)c_{b}(t) + \frac{1}{2}\delta^{a}_{b} \right) \left(\bar{c}^{r}(t^{\prime})c_{s}(t^{\prime}) + \frac{1}{2}\delta^{r}_{s}\right)\right>= \delta^{a}_{b} \delta^{r}_{s}\left[ \Delta(0; \vartheta)-\frac{1}{2}\right]^{2} \nonumber\\ &- \delta^{a}_{s}\delta^{r}_{b}\Delta(t - t'; \vartheta)\Delta(t' - t; \vartheta) 
	= \delta^{a}_{b}\delta^{r}_{s}\left(\frac{i}{2} \tan \frac{\vartheta}{2} - \frac{1}{2}\right)^{2} + \delta^{a}_{s}\delta^{r}_{b}\left(2 \cos \frac{\vartheta}{2}\right)^{-2}.
	\label{colour2}
\end{align}
Putting this together we get a contribution
\begin{align}
&		\frac{1}{4}\int_{\Lambda^{-2}}^{\infty} \frac{dT \,e^{-m^{2}T}}{(4\pi)^{\frac{D}{2}}T^{\frac{D-2}{2}}} \int_{0}^{2\pi} \frac{d \varphi}{2\pi} \int_{0}^{2\pi} \frac{d \vartheta}{2\pi} \, e^{i\vartheta \big(1 - \frac{N^{2}}{2}\big)}e^{i \varphi \big(r - \frac{D}{2}\big)}\left(2\cos\frac{\vartheta}{2}\right)^{N^{2}}\left(2 \cos\frac{\varphi}{2}\right)^{D-2	} \nonumber \\
	&\times\int d^{D}x \left[ d_{aac}d_{rrt}\left(\frac{i}{2} \tan \frac{\vartheta}{2} - \frac{1}{2}\right)^{2} + (d_{abc} d_{abt}+ f_{abc}f_{abt})\left(2\cos \frac{\vartheta}{2}\right)^{-2}  \right] \nonumber \\
	&\times	\int_{0}^{1}dt \int_{0}^{1}dt^{\prime} \,\left<\tr \left(F^{c}(x_\pm) \cdot F^{ t}(x'_\pm)\right)\right>.
	\label{SecF}
\end{align}
To complete the integrals over the modular parameters we need (\ref{Ints1}) with the additional identity
\begin{align}
	\int_{0}^{2\pi} \frac{d \varphi}{2\pi} e^{i\varphi\left(r - \frac{D}{2}\right)}\left(2 \cos \frac{\varphi}{2}\right)^{D - 2n} &= \oint \frac{d\omega}{2\pi i}\omega^{r - (n + 1)}\left(1 + \frac{1}{\omega}\right)^{D - 2n} \nonumber \\
	&= \nCr{D - 2n}{r -n}; \quad D - 2n \geqslant 0, \quad r - n \geqslant 0.
	\label{Ints2}
\end{align}
Armed with these results we first note that
\begin{align}
	\int_{0}^{2\pi}\frac{d \vartheta}{2\pi} \, e^{i\vartheta \big(1 - \frac{N^{2}}{2}\big)}\left(2\cos\frac{\vartheta}{2}\right)^{N^{2}}\left(\frac{i}{2} \tan \frac{\vartheta}{2} - \frac{1}{2}\right)^{2} &= \frac{N^{2} - 4}{4} - \frac{N^{2} - 2}{2} + \frac{N^{2}}{4} \nonumber \\
	&= 0
\end{align}
which kills the first contribution in (\ref{SecF}). The other contribution requires 
\begin{equation}
\begin{split}
	\int_{0}^{2\pi} \frac{d \varphi}{2\pi} \int_{0}^{2\pi} \frac{d \vartheta}{2\pi} \, e^{i\vartheta \big(1 - \frac{N^{2}}{2}\big)}e^{i \varphi \big(r - \frac{D}{2}\big)}\left(2\cos\frac{\vartheta}{2}\right)^{N^{2}-2}\left(2 \cos\frac{\varphi}{2}\right)^{D-2} =& \nCr{D-2}{r-1}\\  =& \begin{cases}1 & \textrm{gauge} \\ 0 & \textrm{ghost}\end{cases}
\end{split}
\end{equation}
so that we finally produce a gauge contribution
\begin{equation}
	\frac{1}{4}\int_{\Lambda^{-2}}^{\infty} \frac{dT \,e^{-m^{2}T}}{(4\pi)^{\frac{D}{2}}T^{\frac{D-2}{2}}}  (d_{abc}d_{abt} + f_{abc}f_{abt}) \int d^{D}x \int_{0}^{1} dt \int_{0}^{1}dt^{\prime} \left<\tr \left( F^{c}(x_\pm) \cdot F_{t}(x'_\pm)\right)\right>.
\end{equation}
Note there is no contribution from the ghost sector with $r = 0$, where no pole appears in  (\ref{Ints2}). This is in agreement with the fact that the field strength tensor does not enter the ghost action (see (\ref{Dets}), for example) -- the auxiliary colour fields correctly project this contribution out of the effective action when $r = 0$ and count it for the gauge field when $r = 1$.

Proceeding with further contributions in the expansion of (\ref{GamExp}) at second order, first note that there can be no cross-term mixing between the $\mathcal{F}$-term and the $\mathcal{V}$-term at this order because self-contractions between the $\bar{\psi}$ and $\psi$ produce traces of $F$. Therefore, we consider two insertions of the $\mathcal{V}$-term  that yield terms quadratic, cubic and quartic in the gauge potential. We are concerned only with the quadratic contributions which take the form
\begin{align}
&	-\frac{1}{4}\int_{\Lambda^{-2}}^{\infty} \frac{dT \,e^{-m^{2}T}}{(4\pi)^{\frac{D}{2}}T^{\frac{D}{2}}} \int_{0}^{2\pi} \frac{d \varphi}{2\pi} \int_{0}^{2\pi} \frac{d \vartheta}{2\pi} \, e^{i\vartheta \big(1 - \frac{N^{2}}{2}\big)}e^{i \varphi \big(r - \frac{D}{2}\big)}\left(2\cos\frac{\vartheta}{2}\right)^{N^{2}}\left(2 \cos\frac{\varphi}{2}\right)^{D} \nonumber \\
	&\times\int d^{D}x \left(i f_{abc} \pm d_{abc}\right)\left(i f_{rst} \pm d_{rst}\right)\int_{0}^{1}dt \int_{0}^{1} dt^{\prime}\,
	\nonumber \\&\times \left<A_{\mu}^{c}(x_\pm)A_{\nu}^{t}(x'_\pm)p^{\mu}(t)p^{\nu}(t')\right> \left< \left(\bar{c}^{a}(t)c_{b}(t) + \frac{1}{2}\delta^{a}_{b} \right) \left(\bar{c}^{r}(t^{\prime})c_{s}(t^{\prime}) + \frac{1}{2}\delta^{r}_{s}\right)\right>.
\end{align}
For the contractions of the colour fields we re-use (\ref{colour2}) and recall that its first term integrates to zero. So we are left with the second contribution which after some shifts of indices becomes
\begin{align}
\hspace{-1em}	\frac{1}{4} &\int_{\Lambda^{-2}}^{\infty} \frac{dT \,e^{-m^{2}T}}{(4\pi)^{\frac{D}{2}}T^{\frac{D}{2}}}  \int_{0}^{2\pi} \frac{d \varphi}{2\pi} \int_{0}^{2\pi} \frac{d \vartheta}{2\pi} \, e^{i\vartheta \big(1 - \frac{N^{2}}{2}\big)}e^{i \varphi \big(r - \frac{D}{2}\big)}\left(2\cos\frac{\vartheta}{2}\right)^{N^{2}-2}\left(2 \cos\frac{\varphi}{2}\right)^{D}\nonumber \\
\hspace{-1em}	&\times\int d^{D}x \left(if_{abc}  \pm d_{abc}\right)\left(if_{abt} \mp d_{abt}\right) \int_{0}^{1}dt \int_{0}^{1}dt^{\prime} \left<A_{\mu}^{c}(x_\pm)A_{\nu}^{t}(x'_\pm)p^{\mu}(t)p^{\nu}(t')\right>
\end{align}
Integrating over the Lorentz fields gives a factor $\nCr{D}{r}$ whilst integrating over the colour fields multiplies this by unity. This leaves
\begin{align}
	-\frac{1}{4}\nCr{D}{r}\int_{\Lambda^{-2}}^{\infty} \frac{dT \,e^{-m^{2}T}}{(4\pi)^{\frac{D}{2}}T^{\frac{D}{2}}} \int d^{D}x & \left(f_{abc}f_{abt}  + d_{abc}d_{abt}\right)\nonumber \\ &\times\int_{0}^{1}dt \int_{0}^{1}dt^{\prime} \left<A_{\mu}^{c+}(t)A_{\nu}^{t+}(t^{\prime})p^{\mu}(t)p^{\nu}(t')\right>\,
\end{align}
where again the cross terms $f_{bac}d_{abd}$ vanish by symmetry. 

In total, then, we have found that the planar (left and right) terms in the expansion of the effective action up to quadratic order in the gauge field that contribute to the UV divergences of the two-point function are, for $r = 0$, $1$,
\begin{align}
\Gamma[A]_{1,r}^\pm	=&-\frac{1}{2}\int_{\Lambda^{-2}}^{\infty} \frac{dT}{T}\frac{ e^{-m^{2}T}}{ (4\pi T)^{\frac{D}{2}}} \int d^{D}x \Bigg\{  \nCr{D}{r} \Bigg[ \int_{0}^{1}dt \Big( \mp \sqrt{T} d_{aac}  \left<A_{\mu}^{c}(x) p_{\mu}\right>\nonumber\\ 
&-\frac{T}{a}\left(f_{acd}f_{ace}+d_{acd}d_{ace}\right)  \left<A_{\mu}^{d}(x)A_{\mu}^{e}(x)\right>  \Big)\nonumber \\
& + \frac{T}{2} \left(f_{abc} f_{abt} + d_{abc}d_{abt}\right) \int_{0}^{1}dt \int_{0}^{1}dt^{\prime} \left<A_{\mu}^{c}(x_\pm)A_{\nu}^{t}(x'_\pm)p^{\mu}p^{\nu}\right> \Bigg] \nonumber \\
& -\frac{1 - (-1)^{r}}{4} T^2 (f_{abc}f_{abt}+d_{abc}d_{abt}) \int_{0}^{1} dt \int_{0}^{1}dt^{\prime} \left<\tr \left( F^{c}(x_\pm) \cdot F_{t}(x'_\pm)\right)\right>\Bigg\}~.
\label{integrales}
\end{align}
Hence,
\begin{align}
\Gamma[A] = \sum_{l=\pm}\Big(\Gamma[A]_{1,1}^l-2 \Gamma[A]_{1,0}^l\Big)
\label{eq:F-effeact}
\end{align}
is the final expression for the (planar part of the) one-loop effective action limited to quadratic order in the gauge field. Whilst this result could have been arrive at from expanding (\ref{eq:gh-eff}) and (\ref{eq:YM-eff}), the unification of the two sectors and the efficient generation of the path ordering would become ever more advantageous for higher order calculations.

\section{Parameter integrals and the $\beta$-function}
\label{secBeta}

In this section we complete the calculation of \eqref{eq:F-effeact} by evaluating the functional integrations in the phase-space variable and the Schwinger proper time integral.

\subsection{One-point function}
\label{sec:one-two-pts}

As mentioned above, for a single insertion, the argument of the gauge field can be shifted. Hence the left and right part of the one-point function are opposite to each other and their sum vanish. In fact, since $\left< A_\mu (x) p_{\mu}\right> =  A_\mu (x) \left<p_{\mu}\right>=0$ those contributions independently vanish.  Although the vanishing of the one-point function  would seem to suggest that the vacuum is stable and ripe to be expanded around, we recall from the Abelian, $U(1)$, case that non-planar diagrams spoil this due to IR divergences at quadratic order in $A$.

\subsection{Two-point function}

Let us consider the contributions to~\eqref{eq:F-effeact} coming from  the first quadratic pieces in~\eqref{integrales}. Since these come from single vertex insertions, the integral over $x\in\mathbb{R}^D$ eliminates the operatorial dependence in the fields, and again the left and right parts are equal. Summing the two sectors according to~\eqref{eq:F-effeact} yields an over factor of $(D-2)$. Hence, 
\begin{align}
\Gamma[A]\supset \frac{D-2}{4}
  \,\frac{1}{(4\pi)^{\frac{D}2}}\,\left(f_{abc}f_{abd}+d_{abc}d_{abd}\right)
  \int_0^\infty \frac{d T }{ T ^{\frac{D}2}}\int_{\mathbb{R}^D}dx_0\ A_\mu^c(x_0)\,A_\mu^d(x_0)
\end{align}
which, using the group identities reported in Appendix~\ref{app1},  reduces to
\begin{align}\label{gh-aaN}
  \frac{N\left(D-2\right)}{2(4\pi)^{\frac{D}2}}
  \,m^{D-2}\,\Gamma(1-\tfrac{D}2,\tfrac{m^2}{\Lambda^2})
  \int_{\mathbb{R}^D}dx\ A_\mu^a(x)\,A_\mu^a(x)
  \,.
\end{align}
Above we have restored the IR- and UV-regulators $m$ and $\Lambda$, respectively. Thus the Schwinger integral is seen to produce a Gamma function. The previous contribution would give a mass to the gauge fields which, because of gauge invariance, is expected to cancel.

Next, let us consider the right-right Moyal contribution of the second quadratic term from~\eqref{integrales}. This contributes to~\eqref{eq:F-effeact} as 
\begin{align}
\label{e1}
&-\frac{D-2}{4(4\pi)^{\frac{D}2}}
(f_{abc}f_{abt}+d_{abc}d_{abt})
\mbox{}\times\nonumber\\
  &\mbox{}\times
  \int_0^\infty \frac{d T}{ T^{\frac{D}2}}
  \int_{\mathbb{R}^D} dx\int_0^1 dt\, dt'
  \left\langle A_\mu^{c}(x_+(t))\,p_\mu(t)\,A_\nu^{t}(x_+(t'))\,p_\nu(t')\right\rangle\nonumber\\[3mm]
&=-\frac{N(D-2)}{2(4\pi)^{\frac{D}2}}
\int_0^\infty \frac{d T}{ T^{\frac{D}2}}
\int d\bar k
  \ \tilde A_\mu^a(k)\,\tilde A_\nu^a(-k)
  \mbox{}\times\nonumber\\
  &\mbox{}\times\int_0^1 dtdt'\ \left\langle p_\mu(t)\,p_\nu(t')
  \ e^{i\sqrt T\,k\,[x(t)-x(t')]-\frac{i}{\sqrt T}\,\theta k\,[p(t)-p(t')]}\right\rangle
\end{align}
If we expand the exponential to order $\mathcal{O}(T)$ in order to determine the UV-divergent part we obtain 
\begin{align}
&-\frac{N(D-2)}{2(4\pi)^{\frac{D}2}}
\int_0^\infty \frac{d T}{ T^{\frac{D}2}}
\int d\bar k
\ \tilde A_\mu^a\,\tilde A_\nu^{a*}\,
\mbox{}\times\nonumber\\
&\mbox{}\times\int_0^1 dtdt'
\left[\frac12\,\delta_{\mu\nu}
  + T\,\delta_{\mu\nu}\,k^2\,G(t-t')
  + T\,k_\mu k_\nu\,\dot G^2(t-t')\right]~.
   \end{align}
  Since the contribution of the left-left part is identical we get (applying $\dot{G}(t)^{2} = \frac{1}{4} + 2G(t)$ and $\int_{0}^{1}dt\,dt'\, G(t-t') = -\frac{1}{12}$)
 \begin{align}\label{gh-apap}
\Gamma[A]\supset &-\frac{N(D-2)}{2(4\pi)^{\frac{D}2}}
\int d\bar k
\ \tilde A_\mu^a(k)\,\tilde A_\nu^{a*}
\mbox{}\times\nonumber\\
&\mbox{}\times\left[m^{D-2}\,\Gamma(1-\tfrac{D}2,\tfrac{m^2}{\Lambda^2})\,\delta_{\mu\nu}
- \frac16\,m^{D-4}\,\Gamma(2-\tfrac{D}2,\tfrac{m^2}{\Lambda^2})
\,\left(\delta_{\mu\nu}k^2-k_\mu k_\nu\right)
\right]\,.
\end{align}
We have used expression \eqref{pp}, \eqref{xx} and \eqref{px} in the computation of the expectation values. In particular, since the mean value $\langle p(t)p(t')\rangle$ does not depend on $t,t'$, then any expectation value involving the combination $p(t)-p(t')$ vanishes (see also Appendix \ref{pbc} for an explicit calculation).
Expression \eqref{gh-apap} shows a mass term which exactly cancels \eqref{gh-aaN}. Moreover, there is a logarithmically divergent term that corresponds to a transversal propagator.

Finally, we consider the last right-right Moyal quadratic contribution from \eqref{integrales},
\begin{align}\label{ff}
  &\frac{1}{4(4\pi)^{\frac{D}2}} (f_{abc}f_{abt} +d_{abc} d_{abt})
  \int_0^\infty \frac{d T}{ T^{\frac{D}2-1}}
  \int dx_0\int_0^1 dtdt'
  \ \left\langle F^{c}_{\mu\nu}(x_+(t))\,F^{t}_{\nu\mu}(x_+(t'))
  \right\rangle\nonumber\\[2mm]
  &=-\frac{N}{2(4\pi)^{\frac{D}2}}
\int_0^\infty \frac{d T}{ T^{\frac{D}2-1}}
\int_0^1 dtdt'\int d\bar k
\ \tilde F^a_{\mu\nu}\,\tilde F^{a*}_{\mu\nu}
\left\langle 
e^{i\sqrt T\,k\,[x(t)-x(t')]-\frac{i}{\sqrt T}\,\theta k\,[p(t)-p(t')]}
\right\rangle\,,
\end{align}
which, expanding the exponential to order $\mathcal{O}(T^0)$, provides divergent contribution
\begin{align}
&-\frac{N}{2(4\pi)^{\frac{D}2}}
\,m^{D-4}\,\Gamma(2-\tfrac{D}2,\tfrac{m^2}{\Lambda^2})
\int dx\ F^a_{\mu\nu}(x)\,F^a_{\mu\nu}(x)~.
\end{align}
Thus, once again adding the other Moyal polarisation, results in the effective action quadratic contribution
\begin{align}\label{ff2}
\Gamma[A]\supset -\frac{4N}{2(4\pi)^{\frac{D}2}}
\,m^{D-4}\,\Gamma(2-\tfrac{D}2,\tfrac{m^2}{\Lambda^2})
\int d\bar k\ \tilde A_\mu^c\,\tilde A_\nu^{c*}
(\delta_{\mu\nu}\,k^2-k_\mu k_\nu)\,.
\end{align}

\subsection{The $\beta$-function}
We have now all the ingredients to obtain the $\beta$-function for $U_\star(N)$ in $D=4$. As is well known the $\beta$ function plays a central role in renormalisation of a quantum field theory and contains the information about the dependence of the couplings on momentum scale. In the particular case of the pure Yang-Mills theory on non-commutative space we are studying here we take into account the divergent one-loop contributions \eqref{gh-aaN}, \eqref{gh-apap} and \eqref{ff2} to obtain for the quadratic effective action, 
\begin{align}
\label{e4}
  &\Gamma_{\rm eff}^{(2)}=\int_{\mathbb{R}^4}d\bar k \
  \tilde A_\mu^a\,\tilde A_\mu^{a*}
  \left(\delta_{\mu\nu}\,k^2-k_\mu k_\nu\right)
  \ \left\{\frac{1}{2g^2}-\frac{1}{32\pi^2}\,
  \left(-\frac{N}3+4N\right)
  \,\Gamma(0,\tfrac{m^2}{\Lambda^2})
  \right\}\nonumber\\[2mm]
  &=\frac{1}{2g^2}
  \left\{1-\frac{11}{48\pi^2}\, g^2N\,\log{(\tfrac{\Lambda^2}{m^2})}\right\}
  \int_{\mathbb{R}^4}dx\
  A_\mu^a
  \left(-\delta_{\mu\nu}\,\partial^2+\partial_\mu\partial_\nu\right)
  A_\nu^a\, .
\end{align}
The one-loop contribution to the $\beta$ function is determined from the running of the coupling constant, $g$, with energy-momentum scale that is implied by (\ref{e4}). Interpreting the coefficient to the space-time integral as defining the coupling at momentum scale $\Lambda$ we get
\begin{equation}
\beta(g)\big\vert_{U_\star(N)}:=
\Lambda\,\partial_\Lambda g=-\frac{11N}{48\pi^2}g^3\,,
\end{equation}
that is in accordance with \cite{Krajewski:1999ja} (see also \cite{Armoni:2000xr}). For the particular $U_\star(1)$ case, the non-standard normalisation of the generator requires $g\to \sqrt2 g$, which gives
\begin{equation}
\beta(g)\big\vert_{U(1)^\star}=-\frac{11N}{24\pi^2}g^3\,,
\end{equation}
first computed in \cite{Krajewski:1999ja} and \cite{Martin:1999aq}.

\section{Discussion}
\label{discussion}
We have proposed a new worldline approach to the effective actions of the $U(N)$ non-commutative gauge theories. As for their Abelian counterparts, these theories are linked to phase-space worldline path integrals. The main new ingredients we present here are sets of auxiliary fields that allow a suitable treatment of color degrees of freedom and the non-trivial Lorentz structure without the need to introduce a path ordering, that also retains manifest gauge-invariance throughout calculations. Moreover, such auxiliary fields allow the unification of the computation of the gauge part and ghost part of the effective action in a single worldline model, distinguished only by the choice of integer occupation numbers corresponding to the different sectors. This has obvious advantages since, as we have illustrated here, one may calculate path integrals for the two sectors simultaneously, whilst working in phase-space avoids many of the familiar ordering ambiguities of operators in non-commutative space. 

In this paper we have focussed on calculating the $\beta$-function, which requires expansion of our path integrals to quadratic order, finding agreement with existing literature. However our path integral is general, in that higher order contributions could be determined with the same method (for example, the $\beta$-function of $U(1)$ theory on non-commutative space time was computed from the two point function in \cite{NCU1} where it was then verified by examination of the three and four point functions). The only difference would be that expectation values of products of a greater number of auxiliary fields would appear, but the path ordered trace would still be produced by their contractions. We also wish to highlight that the path integral (\ref{GamExp}) can be immediately adapted to calculate one-loop gluon scattering amplitudes by specialising the gauge potential to a sum over plane waves; this ongoing work is a further application of our approach.

Moreover, although the divergences relevant for the determination of the $\beta$-function required examination only of the planar contributions (involving products of the gauge field evaluated at $x_{+}$ only or at $x_{-}$ only), our worldline theory includes also the non-planar contributions that could be calculated easily within our framework. These would include the twisting term caused by the non-commutativity parameters that enter the exponential of our insertions. The extension to irreducible representations of $U(N)$ with mixed symmetry is also immediate with the same technique as reported in \cite{JO1, JO2}. Given the recent analysis of \cite{PropNC} it would be worth investigating the worldline representation of the propagator (requiring open worldlines) of both matter fields and photons or gluons using the phase space techniques we have presented here, or to include an additional external field that is well known to be accessible using first quantised techniques. 

As is well-known, renormalisation of field theories on Moyal spacetime could be spoiled by UV/IR mixing, a new type of IR singularity introduced by the high-energy virtual particles of non-planar diagrams. Two-loop calculations on the noncommutative torus show that compactness could tame IR singularities \cite{DAscanio:2016jmt}. A successful mechanism discovered by Grosse and Wulkenhaar \cite{Grosse:2003nw} has lead to the formulation of renormalisable field theories on Moyal spacetime through the introduction of an appropriate (harmonic oscillator) background. A similar mechanism on a scalar field under a magnetic background provides an exactly solvable but translation invariant model \cite{Langmann:2003cg,Langmann:2003if}. Although the Grosse-Wulkenhaar (GW) term is difficult to implement in gauge theories, a very interesting candidate for a renormalisable NC gauge theory, known as induced gauge theory \cite{deGoursac:2007fzf,Grosse:2007dm}, introduces the GW term through covariant coordinates \cite{Madore:2000en} so that gauge invariance is preserved. Also, an appealing extension of this model -- and with a trivial vacuum -- has been formulated in terms of a more general class of covariant coordinates and an appropriate gradation of Moyal algebra \cite{deGoursac:2008bd}. We expect our results can help in the involved perturbative study of these and similar models (see e.g.\ \cite{Blaschke:2009aw}).

We believe that this worldline theory could also be useful for future work incorporating gravitational interactions (as have previously been studied using first quantised techniques \cite{Grav1, Grav2}) or to study higher loop-order corrections to the $\beta$-function using existing worldline methods proven to have certain advantages over the standard formalism. In particular, since our worldline theory is valid for off-shell amplitudes, it can be used to form higher order diagrams by ``sewing'' pairs of gauge particles (or by incorporating virtual gluons into a modification of the worldline Green functions \cite{Multi}). Moreover, an obvious extension to this work (or even the $U_{\star}(1)$ version) would be to include matter fields coupled to the Yang-Mills fields and to study the modifications to the one-loop effective action due to such interactions in the spirit of Euler-Heisenberg effective actions. One more natural direction of this series of work would be the study of Schwinger electron-positron pair production by a constant external electric field in non-commutative QED which has been studied in \cite{Chair:2000vb}; for a review on strong field and non-commutative QED see \cite{Ilderton:2010rx}.

\section*{Acknowledgements}
The authors are pleased to thank Christian Schubert for useful discussions and comments on the manuscript. The work of NA was supported by IBS (Institute for Basic Science) under grant No. IBS-R012-D1. JPE thanks PRODEP (Project 671/PRODEP/2017) for support during this work and Anabel Trejo for helpful conversations. The work of PP is supported by CONICET.

\appendix
\section{Some group identities}
\label{app1}
It is useful to note the following identities involving the structure constants for different choices of Lie group. For $U_\star(N)$ on non-commutative space we find
\begin{equation}
f_{abc}f_{abd} + d_{abc}d_{abd} = \textrm{diag}\left(2N, 2N,2N,2N, \ldots ,2N\right) = 2N \delta_{cd}.
\end{equation}
In a commutative space the equivalent results for $U(N)$ and $SU(N)$ do not involve the symmetric structure constants, so we find (the indices range over integers $\{0,1, \ldots, N\}$ for $U(N)$ and over $\{1, 2, \ldots\, N\}$ for $SU(N)$)
\begin{equation}
	f_{abc}f_{abd} = 
	\begin{cases}
		 \textrm{diag}\left(N, N, \ldots, N\right) & SU(N)\\
		 \textrm{diag}\left(0, N, N, \ldots, N\right) & U(N) \\
	\end{cases} 
\end{equation}
Finally, although $SU_\star(N)$ is not a consistent Yang-Mills theory, it is still possible to compute the symmetric structure constants for this group and for comparison we record the result (indices range over integers $\{1, 2, \ldots, N\}$)
\begin{equation}
	f_{abc}f_{abd} + d_{abc}d_{abd}=	 \textrm{diag}\left(\frac{2(N^{2} - 2)}{N},\frac{2(N^{2} - 2)}{N}, \ldots, \frac{2(N^{2} - 2)}{N} \right).
\end{equation}
It is useful to note that for the unitary group the sum decomposes (the first part represents the commutative result) as:
\begin{align}
	f_{abc}f_{abd} &= \textrm{diag}\left(0, N, N, \ldots, N\right) \\
	d_{abc}d_{abd} &= \textrm{diag}\left(2N,N,N, \ldots N\right).
\end{align}
We also have $(d_{cab}+if_{cab})(d_{dba}+if_{dba}) = f_{abc}f_{abd} + d_{abc}d_{abd}$ due to cancellation of the cross terms.

\section{Worldline Green functions}\label{pbc}

We derive the worldline Green functions associated to the phase-space degrees of freedom $x$ and $p$ by considering the generating functional of 
\begin{align}\label{mv}
  \mathcal{Z}[k, j] := \left\langle\,e^{i\int_0^1 dt\left\{k(t)p(t)+j(t)x(t)\right\}}\,\right\rangle
  =\frac{ \int\mathscr{D}x(t)\mathscr{D}p(t)\ e^{-\int_0^1 dt\,\left\{p^2-ip\dot{x}\right\}}
  \,e^{i\int_0^1 dt\left\{kp+jx\right\}}}
  {\int\mathscr{D}x(t)\mathscr{D}p(t)\ e^{-\int_0^1 dt\,\left\{p^2-ip\dot{x}\right\}}}\,,
\end{align}
with periodic boundary conditions on the particle trajectories $x(t)$ and without restriction on their momenta, $p(t)$; however, since these conditions involve a zero mode, we will integrate over the subspace of trajectories which are orthogonal to this zero mode.

Let us first write expression \eqref{mv} as
\begin{align}\label{mvm}
  \left\langle\,e^{i\int_0^1 dt\left\{k(t)p(t)+j(t)x(t)\right\}}\,\right\rangle
  =\frac{\int\mathscr{D}Z(t)\ e^{-\frac12\int_0^1 dt\,Z(t)^T\,D\,Z(t)}\,e^{i\int_0^1 dt\,Z(t)^T J(t)}}
  {\int\mathscr{D}Z(t)\ e^{-\int_0^1 dt\,Z(t)^T\,D\,Z(t)}}\,.
\end{align}
where $Z(t)$ denote trajectories in phase-space and $J(t)$ the corresponding external sources,
\begin{align}
  Z(t)=\left(\begin{array}{c}p(t)\\x(t)\end{array}\right)\,,\qquad
  J(t)=\left(\begin{array}{c}k(t)\\j(t)\end{array}\right)\,.
\end{align}
We have also defined the matricial operator
\begin{align}
  D=\left(\begin{array}{cc}2&-i\partial_t\\[2mm]
  i\partial_t&0\end{array}\right)\,.
\end{align}
For periodic boundary conditions $D$ is symmetric but has a zero mode
\begin{align}
  Z_0(t)=\left(\begin{array}{c}0\\1\end{array}\right)
\end{align}
and is therefore not invertible. However, expanding about the loop center of mass, $x(t) \rightarrow x_0+ q(t)$, the deviation $q(t)$ must be periodic and integrate to zero (string inspired boundary conditions, $q(t) = q(0)$ and $\int_{0}^{1} dt\, x(t) = 0$). Then the inverse in this subspace orthogonal to $Z_0(t)$ is given by
\begin{align}
  D^{-1}=\left(\begin{array}{ccc}
  \frac12
  &&-\frac{i}{2}\epsilon(t-t')+i(t-t')\\[2mm]
  \frac{i}{2}\epsilon(t-t')-i(t-t')
  &&-|t-t'|+(t-t')^2+\frac16
  \end{array}\right)\,.
\end{align}
We can now complete the square in expression \eqref{mvm} and express the expectation value in terms of the inverse operator $D^{-1}$. The result reads
\begin{align}\label{meanvalueapp}
  &\left\langle\,e^{i\int_0^1 dt\left\{k(t)p(t)+j(t)x(t)\right\}}\,\right\rangle
  =e^{-\frac12\int_0^1 dt\,J(t)^T\,D^{-1}\,J(t)}
  \nonumber\\
  &=\exp{\left(-\int\!\int dtdt'\left\{\tfrac14\,k(t)k(t')+g(t,t')j(t)j(t')
  +\tfrac{i}{2}\, h(t,t')k(t)j(t')\right\}\right)}
  \,,
\end{align}
with
\begin{align}
  g(t,t')&:=-\frac12|t-t'|+\frac12 (t-t')^2+\frac1{12}\,,\\
  h(t,t')&:=2\partial_t g(t,t') = -\epsilon(t-t')+2(t-t')\,.
\end{align}
When employing the worldline formalism to compute one-loop gluon amplitudes, the source reads $j^\mu(t) \sim\sum_i k_i^\mu \delta(t-t_i)$, where $k_i$ are the gluon momenta. Since the latter is coupled to $x^\mu_0+q^\mu(t)$, the integral over the zero mode $x_0$ enforces the overall momentum conservation, and in turn the current $j(t)$ satisfies 
\begin{align}\label{q=0}
    \int_0^1 dt\,j(t)=0\,.
\end{align}
Since under condition~\eqref{q=0} terms in $g(t,t')$ which depend only on $t$ or on $t'$ (but not on both), as well as terms in $h(t,t')$ which do not depend on $t'$, are irrelevant in~\eqref{meanvalueapp}, we can instead use the simplified Green functions:
\begin{align}
  G(t-t')&:=-\frac12|t-t'|+\frac12(t-t')^2\,,\\
  H(t-t')&:= 2\dot G(t-t')=-\epsilon(t-t')+2(t-t')\,,
\end{align}
that are homogeneous string-inspired Green functions. From here, it is easy to derive the Wick contractions by functional differentiation with respect to the sources at fixed times:
\begin{equation}
\begin{split}
  \langle p_\mu(t) p_\nu(t')\rangle&=\frac12\,\delta_{\mu\nu}\,,\\
  \langle x_\mu(t) x_\nu(t')\rangle&=2\,\delta_{\mu\nu}\,G(t-t')\,,\\
  \langle p_\mu(t) x_\nu(t')\rangle&=i\,\delta_{\mu\nu}\,\dot G(t-t')\,,
  \end{split}
  \label{eq:x-p-correlator}
\end{equation}
as reported in the main text. Moreover, the correlator~\eqref{meanvalueapp} becomes
\begin{align}\label{meanvalueapp-f}
  &\left\langle\,e^{i\int_0^1 dt\left\{k(t)p(t)+j(t)x(t)\right\}}\,\right\rangle
  \nonumber\\
  &=\exp{\left(-\int\!\int dtdt'\left\{\tfrac14\,k(t)k(t')+G(t-t')j(t)j(t')
  +i \dot G(t-t')k(t)j(t')\right\}\right)}
  \,.
\end{align}
Insertions of $x$ and $p$ can now be generated by derivation with respect to $j$ and $k$.

As far as the auxiliary field are concerned, the same process can be used to derive the two-point functions. However it is easy to check that the Green function of $\partial_t$ consistent with boundary conditions~\eqref{eq:BCs} is
\begin{align*}
\Delta(t-s;\varphi)= \frac{1}{2\cos\frac{\varphi}{2}}\Bigl[ e^{i\frac{\varphi}{2}} \Theta(t-s) - e^{-i\frac{\varphi}{2}} \Theta(s-t)\Bigr]
\end{align*}
which satisfies the following properties
\begin{align}
& \Delta(0;\varphi) = \frac i2\tan\frac{\varphi}{2}\nonumber\\
& \Delta(x;\varphi) \Delta(-x;\varphi) = -\frac{1}{\left(2\cos\frac{\varphi}{2}\right)^2}
\end{align}
that are employed several times in the manuscript. The generating functional of the fields $\bar{c}$ and $c$ is
\begin{equation}
	\left< \e^{-\int dt\, \{\bar{l}^{a}c_{a} + \bar{c}^{a}l_{a} \} } \right> = \exp\left(\int\!\int dt ds\, \bar{l}^{a}(t) \Delta(t-s; \varphi) l_{a}(s)\right)
\end{equation} 
from which insertions of $\bar{c}$ and $c$ can be produced by differentiating with respect to (Grassmann) sources $l$ and $\bar{l}$. Here we have used that $\Delta(t-s; \varphi)^{\star} = \Delta(s-t; -\varphi) = - \Delta(t-s; \varphi)$.

To terminate this section let us compute the correlator
\begin{align}
C_{\mu\nu} (t_1,t_2):=\left<p_{\mu} (t_1)p_{\nu} (t_2)e^{i\sqrt T\,k\,[x(t_1)-x(t_2)]-\frac{i}{\sqrt T}\,\theta k\,[p(t_1)-p(t_2)]}\right>~,
\end{align}
 that appears in the evaluation of the quadratic part of the effective action (c.f. (\ref{e1})). Firstly, notice that
\begin{align}
\big\langle p_\mu (t_i) (p_\alpha(t_1) -p_\beta(t_2))\big\rangle =0
\end{align} 
as the $p$ correlator is $t$-independent. Thus,
\begin{align}
C_{\mu\nu} (t_1,t_2):=&
\left\lbrace \frac12 \delta_{\mu\nu} -Tk_\alpha k_\beta 
\Big\langle p_\mu(t_1) (x^\alpha (t_1)-x^\alpha(t_2))\Big\rangle \Big\langle p_\nu(t_2) (x^\beta (t_1)-x^\beta(t_2))\Big\rangle \right\rbrace \nonumber\\
&\times \Big\langle e^{i\sqrt T\,k\,[x(t_1)-x(t_2)]-\frac{i}{\sqrt T}\,\theta k\,[p(t_1)-p(t_2)]}\Big\rangle\nonumber\\
=& \left\lbrace \frac12 \delta_{\mu\nu}+ Tk^2 \dot G^2(t_1-t_2) \right\rbrace 
\Big\langle e^{i\sqrt T\,k\,[x(t_1)-x(t_2)]-\frac{i}{\sqrt T}\,\theta k\,[p(t_1)-p(t_2)]}\Big\rangle
\end{align}
where, in the last equality we have used the correlators~\eqref{eq:x-p-correlator}. The leftover expression can be easily obtained by plugging into the generic formula~\eqref{meanvalueapp-f} the suitable currents
\begin{equation}
\begin{split}
& j^\mu(t) =\sqrt{T} k^\mu \big(\delta(t-t_1) -\delta(t-t_2)\big) \\
 & k^\mu(t) =-\frac{1}{\sqrt{T}} (\theta k)^\mu \big(\delta(t-t_1) -\delta(t-t_2)\big)~,
  \end{split}
  \label{eq:currents}
\end{equation} 
which finally yield
 \begin{align}
 C_{\mu\nu} (t_1,t_2)=\left\lbrace \frac12 \delta_{\mu\nu} +Tk^2 \dot G^2(t_1-t_2) \right\rbrace e^{2T k^2 G(t_1-t_2)}~.
 \end{align}
 Note in particular, that the $\theta$-dependent part of the correlator identically vanishes, thus this two-point function provides a fully planar contribution. On the other hand, a mixed  correlator
 $\left\langle A_\mu (x_+) p^\mu (t) A_\nu (x_-') p^\nu(t') \right\rangle$  would have a $\theta$-dependent (twist) part in the exponential, since the delta functions in the current $k^\mu(t)$ of~\eqref{eq:currents} would appear with a relative positive sign, and would thus contribute to the non-planar part of the effective action.

\bibliographystyle{JHEP}
\bibliography{bibNCYM}

\end{document}